\documentclass[useAMS,usenatbib]{mn2e}
\usepackage{graphics,graphicx}
\usepackage{float}

\title[The Elephant Trunk Nebula  and the Trumpler~37 cluster]{The Elephant Trunk Nebula  and the Trumpler~37 cluster: Contribution of triggered star formation to the total population of an H\,{\sc{ii}} region}

\author[K. V. Getman et al.]{Konstantin V. Getman$^{1}$\thanks{E-mail: gkosta@astro.psu.edu (KVG)}, Eric D. Feigelson$^{1,2}$, Aurora Sicilia-Aguilar$^{3}$,\newauthor Patrick S. Broos$^{1}$, Michael A. Kuhn$^{1}$, Gordon P. Garmire$^{1}$\\
$^{1}$Department of Astronomy \& Astrophysics, 525 Davey Laboratory, The Pennsylvania State University, University Park, PA 16802, USA\\
$^{2}$Center for Exoplanets and Habitable Worlds, The Pennsylvania State University, University Park, PA 16802, USA\\ 
$^{3}$Departamento de Fisica Terica, Universidad Aut noma de Madrid, Cantoblanco 28049, Madrid, Spain}

\begin{document}

\date{Accepted for publication in MNRAS, 2012 August 6.}

\pagerange{\pageref{firstpage}--\pageref{lastpage}} \pubyear{2012}

\maketitle

\label{firstpage}

\begin{abstract}
Rich young stellar clusters produce H\,{\sc{ii}} regions whose expansion into the nearby molecular cloud is thought to trigger the formation of new stars. However, the importance of this mode of star formation is uncertain. This investigation seeks to quantify triggered star formation (TSF) in IC 1396A (a.k.a., the Elephant Trunk Nebula), a bright rimmed cloud (BRC) on the periphery of the nearby giant H\,{\sc{ii}} region IC 1396 produced by the Trumpler 37 cluster. X-ray selection of young stars from {\it Chandra} X-ray Observatory data is combined with existing optical and infrared surveys to give a more complete census of the TSF population. Over 250 young stars in and around IC 1396A are identified; this doubles the previously known population.  A spatio-temporal gradient of stars from the IC 1396A cloud toward the primary ionizing star HD 206267 is found.  We argue that the TSF mechanism in IC 1396A is the radiation-driven implosion process persisting over several million years. Analysis of the X-ray luminosity and initial mass functions indicates that $>140$ stars down to 0.1~M$_{\odot}$ were formed by TSF. Considering other BRCs in the IC 1396 H\,{\sc{ii}} region, we estimate the TSF contribution for the entire H\,{\sc{ii}} region exceeds $14-25$\% today, and may be higher over the lifetime of the H\,{\sc{ii}} region. Such triggering on the periphery of H\,{\sc{ii}} regions may be a significant mode of star formation in the Galaxy. 
\end{abstract}

\begin{keywords}
ISM: individual (Elephant Trunk Nebula, IC 1396A) - ISM: clouds - open clusters and associations: individual (IC 1396, Trumpler 37) - planetary systems: protoplanetary disks - stars: formation - stars: pre-main sequence - X-rays: stars
\end{keywords}

\section{INTRODUCTION}\label{introduction_section}

\subsection{Motivation}\label{motivation_subsection}
A long-standing issue in Galactic star formation is: How much star formation relies on spontaneous gravitational collapse of cold clouds? $vs.$ How much star formation relies on collapse ``triggered'' by the imposition of external forces? Triggered star formation (TSF) may play roles on large-scales from compression by Galactic spiral arms, on meso-scales from compression by supernova remnant (SNR) superbubbles in starburst complexes, and on small-scales from compression by expanding H\,{\sc{ii}} regions \citep{Elmegreen07}. It has long been recognized that star formation in molecular clouds can be triggered by ionization or shock fronts produced by nearby massive stars \citep{Elmegreen77}. Two major mechanisms for triggering by expanding H\,{\sc{ii}} regions have been widely discussed: the Radiation Driven Implosion (RDI) model and the Collect and Collapse (C\&C) model. 

In the RDI model, the expanding OB ionization front ablates the surface of surrounding pre-existing cloudlets producing bright-rimmed clouds (BRCs), often with cometary structures, and driving inward a compressional shock that induces star formation. Discussed since the 1980s, RDI now has well-developed hydrodynamical calculations showing the viability of TSF on timescales $\sim 0.1$~Myr \citep[][and references therein]{Kessel03,Miao09,Bisbas11}.

Observationally, two types of bright-rimmed clouds (BRCs) can be distinguished: relatively massive cloudlets with widths of $w>0.5-0.7$~pc, which are often found to harbor more than a few stars at their tips; and thin ($w<0.5$~pc) elongated pillar-like structures with a few or no stars at their tips \citep{Chauhan11}. According to the RDI model, this distinction may reflect different evolutionary stages of BRCs with narrow pillars interpreted as the last vestiges of the photoevaporated and dispersed cloudlet that has insufficient amount of remaining gas to form new stars \citep{Bisbas11}. Alternative explanations for the thin elongated pillars involve hydrodynamic instabilities in ionization fronts \citep{Mizuta06,Whalen08} or ionization of large turbulent clouds without pre-existing cloudlets \citep[][and references therein]{Gritschneder10}.

In most cases, molecular, infrared (IR), and H$\alpha$ surveys of BRCs trace only the most recently formed stars \citep[e.g.,][]{Sugitani95,Ogura02,Urquhart09}. While longer wavelength surveys select principally the youngest stars undergoing accretion from dusty disks, X-ray surveys efficiently select young stars due to enhanced magnetic reconnection flaring that extends to the main sequence \citep{Feigelson10}. In a few cases, $Chandra$ observations have added the diskfree pre-main sequence (PMS) populations to TSF clouds \citep[e.g.,][]{Getman07, Getman09}. Triggered BRCs often show embedded mid-IR sources denoting protostars, while H$\alpha$, $JHK$, and X-ray surveys reveal small clusters of disky and diskfree PMS stars within and in front of the bright rim\footnote{Henceforth, the term ``in front of'' refers to the region between a globule and an ionizing star(s) in 2-dimensional projection, and not the region between the observer and the globule.}. In a few cases, spatial-age gradients in the stellar population are clearly seen where the youngest stars are embedded and older stars are aligned toward the ionizing sources \citep[e.g.,][]{Ogura07, Getman07, Ikeda08, Getman09}.  This directly supports the RDI mechanism and implies that the existing clouds have been actively forming stars for more than 1~Myr.

In the C\&C model, ionization and wind shock fronts sweep up neutral material into a dense shell that becomes gravitationally unstable and fragments. Individual stars or entire massive clusters may form inside these fragments. The theory is moderately developed \citep{Whitworth94, Hosokawa06, Dale07} and may explain a number of spectacular examples of embedded clusters found around H\,{\sc{ii}} regions using long-wavelength (mm to near-IR) observations \citep{Deharveng03, Zavagno06, Zavagno07}. They find a central ionizing star with a roughly spherical H\,{\sc{ii}} region surrounded by a shell of dense molecular gas with dense fragments harboring young stellar objects. However, recent SPH simulations by \citet{Walch11} show that if an H\,{\sc{ii}} region expands into a clumpy molecular cloud then the production of a molecular shell with dense massive fragments does not require the C\&C mechanism. Instead, triggered star formation would take place through simultaneous enhancement of density and global radiation-driven implosion of pre-existing molecular clumps.

Star formation triggered by H\,{\sc{ii}} regions may be very common. The {\it Spitzer}/GLIMPSE survey finds $\sim$10 mid-IR bubbles/deg$^{2}$ throughout the Galactic Disk \citep{Churchwell06, Simpson12}. \citet{Deharveng10} show that most of these bubbles enclose H\,{\sc{ii}} regions ionized by O-B2 stars, many are surrounded by cold dust shells and many have bright-rimmed dust condensations protruding inside the H\,{\sc{ii}} region. $26$\% of the dust shell condensations have ultracompact H\,{\sc{ii}} (UCH\,{\sc{ii}}s) regions and/or methanol masers, likely indicating triggering of massive star formation. Based on the cross-correlation of the positions of the mid-IR shells with the positions of the very bright and red young stellar objects (YSOs) from the Red MSX Source (RMS) catalog, \citet{Thompson11} estimate that the formation of $14-30$\% of massive stars in the Milky Way could be triggered by expanding H\,{\sc{ii}} regions.

\begin{figure*}
\centering
\includegraphics[angle=0.,width=170mm]{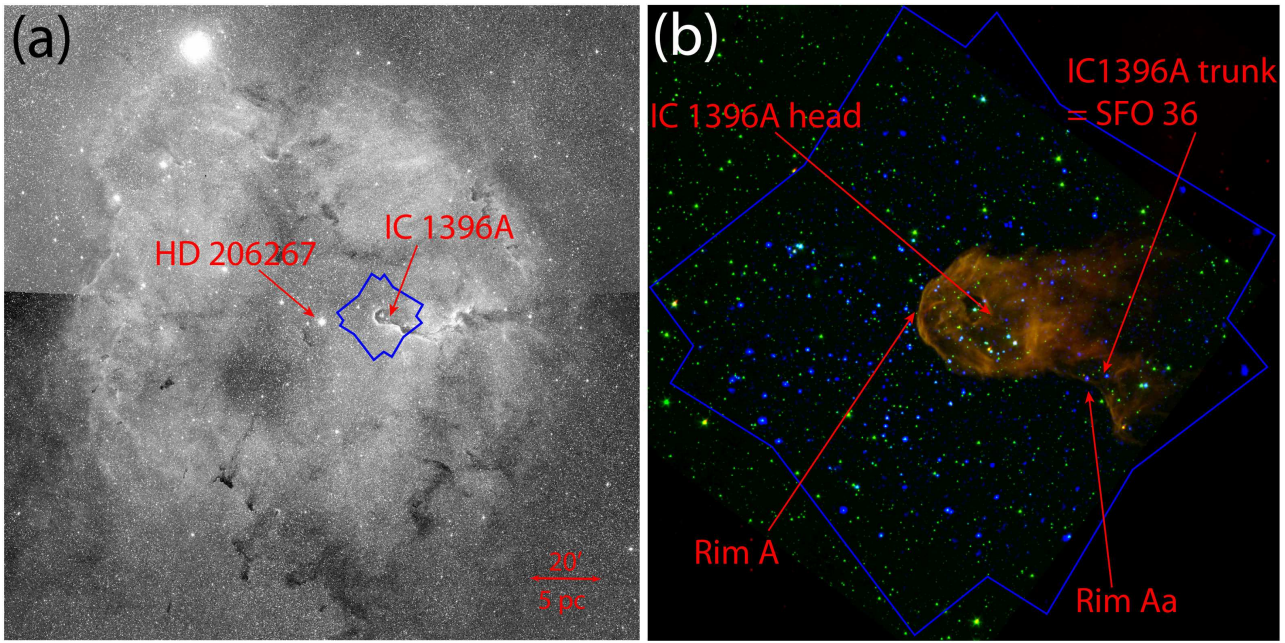}
\caption{(a) A $3\degr \times 3\degr$ image of the emission nebula IC 1396 from the Digitized Sky Survey (DSS). The primary ionizing source of the region, the O6 star HD 206267, and the bright-rimmed globule IC 1396 are marked in red. The blue polygon outlines the ACIS field centered on the IC 1396A. (b) Close-up, combined X-ray and IR image of IC 1396A. The adaptively smoothed {\it Chandra}-ACIS image in the $0.5-8.0$~keV band (blue) is superposed on the $Spitzer$-IRAC composite image in the 3.6~$\mu$m (green) and 8.0~$\mu$m (red) bands. The  ACIS field is outlined in blue; the head and the trunk of the cloud are marked in red. The positions of the two brightest features of the ionized optical rim in front of the cloud are marked in red as Rims A and Aa. North is up, east is to the left \label{fig_intro}}
\end{figure*}

Due to violent energy feedback from massive stars in large star-forming complexes and individual H\,{\sc{ii}} regions, the geometry, conditions, and history of star formation are often difficult to ascertain. The principal signatures of small-scale TSF --- a star cluster in and around the cloudlet on the periphery of an expanding HII region --- may be erased through removal of the cloudlet by stellar winds and UV radiation, or the kinematic dispersal of the unbound cluster within less than a few million years. Thus, it is often difficult to identify sites of triggered star formation and to quantify the impact of triggering processes. IC 1396 is a nearby \citep[$D \sim 870$~pc;][]{Contreras02}, large ($\sim 12$~pc in radius) shell-like H\,{\sc{ii}} region, where traces of recent triggered star formation are still evident \citep[][and references therein]{Kun08}. The goal of the current study is to identify, understand, and quantify the triggered stellar population in the central part of the region associated with the largest BRC, IC 1396A (The Elephant Trunk Nebula), and to estimate the contribution of triggered star formation to the total stellar population in both the central part and, with some additional assumptions, in the entire H\,{\sc{ii}} region.

\subsection{The target: IC~1396A globule}\label{ic1396_subsection}

The IC 1396 H\,{\sc{ii}} region, Sh 2-131 at $(l,b)=(99.3^\circ,3.7^\circ)$, is excited mainly by the O6.5f star HD 206267 located at the center of the region in the 4~Myr old cluster Tr~37 \citep{Sicilia-Aguilar05}[hereafter SA05]. IC 1396 has a rich population of $>20$ bright-rimmed and cometary globules seen in silhouette against the emission nebula \citep[][and Figure \ref{fig_intro} here]{Sugitani91, Froebrich05}. Many of the clouds reside on the large molecular shell surrounding the H\,{\sc{ii}} region, which expands at a speed of $5$~km~s$^{-1}$ with an inferred expansion time around $2.5$~Myr \citep{Patel95}. Sites of possible star-formation have been identified in/around at least several globules \citep[e.g.][]{Schwartz91, Ogura02, Froebrich05} including sites of substantial star-formation in IC~1396N \citep{Nisini01, Getman07, Beltran09, Choudhury10}, SFO~37 \citep{Ikeda08}, and IC~1396A \citep{Reach04, Sicilia-Aguilar05, Sicilia-Aguilar06, Morales-Calderon09, Barentsen11}.

Lying $\sim 15\arcmin$ ($\sim 3.7$~pc projected distance) west of HD~206267, IC 1396A is the bright-rimmed cloud closest on the sky to HD~206267. Its optical rim is the brightest of all rims and the cloud is thus likely to be the physically closest to HD 206267 \citep{Weikard96}. Following \citet{Weikard96} we designate the two brightest features of the rim as Rim A located in front of the $>1$~pc diameter head of the globule and Rim Aa located further west at the shoulder (trunk) of the cloud (Figure \ref{fig_intro}). The part of the cloud behind Rim Aa has its own designation: SFO~36 \citep{Sugitani91}. A 0.3~pc diameter cavity in the molecular cloud is present in the head of the globule produced by the Herbig Ae star LkHa~349 \citep{Hessman95}. Outside this ``hole'', the globule is optically thick ($A_V \sim 10$~mag) with molecular mass $\sim 200$~M$_{\odot}$ \citep{Patel95, Weikard96}.

Early {\it Spitzer} images revealed both bright diffuse mid-infrared (MIR) emission and 17 Class~II/I stars in the IC 1396A globule \citep[][their Figure 2]{Reach04}.  A spectacular multi-band movie of the cloud was one of the first press releases of the {\it Spitzer} mission\footnote{The press release can be found at http://www.spitzer.caltech.edu/images/1058-ssc2003-06b-Dark-Globule-in-IC-1396}. SA05 and \citet{Sicilia-Aguilar06} combined new deeper and wider-area {\it Spitzer} photometry with optical photometry and spectroscopy to reveal a rich lightly obscured population of $>200$ H$\alpha$ emission and/or Li~6707~\AA\ absorption stars. The population is a mixture of $\sim 4$~Myr old members of the central Tr~37 cluster and $\sim 1$~Myr stars spatially concentrated in an arc in front of the globule (see Figure 11 in SA05). In addition, $\sim 50$ heavily obscured IR-excess-selected objects were found inside the cloud \citep{Sicilia-Aguilar06}. Deeper {\it Spitzer}-IRAC maps from the time-series monitoring of the globule \citep{Morales-Calderon09} [hereafter MC09] uncovered an additional dozen heavily embedded Class I/II YSOs. The spatial distribution of sixty five disky members in/around the globule is shown in Figure 16 of MC09. Recently, using the data from the INT Photometric H-Alpha Survey (IPHAS), \citet{Barentsen11} [hereafter B11] identified about two dozen low-mass high-accretion T-Tauri systems spatially concentrated in an arc in front of the globule (see Figure 18a in B11). About half of these were newly discovered stellar members of the region. Spatial clustering of optical stars with younger ages and increasing accretion rates away from the ionizing star HD 206267 (SA05, B11) led SA05 and B11 to propose the presence of a stellar population in front of the IC 1396A globule whose formation has been triggered by HD 206267. Thus, several high-sensitivity optical and mid-infrared surveys have revealed a few hundred young stars in the Tr 37 region, with the youngest concentrated within and around IC 1396A.

\begin{table*}\scriptsize
\centering
 \begin{minipage}{180mm}
 \caption{Basic X-ray Source Properties. This table is available in its entirety in the electronic edition of the journal. A portion is shown here for guidance regarding its form and content. Source net counts, background counts, median energy, and apparent photometric flux are given for the full $(0.5-8)$~keV band. Column 1: X-ray source number. Column 2: IAU designation. Columns 3-4: Right ascension and declination for epoch J2000.0 in degrees. Column 5: 1~$\sigma$ error circle around source position. Column 6: Average off-axis angle for the two merged X-ray observations. Column 7: Source net counts in merged apertures and their 1~$\sigma$ upper errors. Column 8: Observed background counts in merged apertures. Column 9: Average PSF fraction for merged observations (at 1.5~keV) enclosed within source aperture. Column 10: Smallest of the three $p$-values (for the full, soft, and hard energy bands) for no-source hypothesis. Column 11: Effective exposure time. Columns 12: Smallest $p$-value for the one-sample Kolmogorov-Smirnov statistic under the no-variability null hypothesis within a single-observation. Column 13: $p$-value for the one-sample Kolmogorov-Smirnov statistic under the no-variability null hypothesis over the combined observations. Columns 14-15: X-ray median energy and apparent photometric flux.}
 \label{tbl_xray_photometry}
 \begin{tabular}{@{\vline }c@{ \vline }c@{ \vline }c@{ \vline }c@{ \vline }c@{ \vline }c@{ \vline }c@{ \vline }c@{ \vline }c@{ \vline }c@{ \vline }c@{ \vline }c@{ \vline }c@{ \vline }c@{ \vline }c@{ \vline }}
\cline{1-15}
&&&&&&&&&&&&&&\\ 
No. & CXOIC1396A & R.A. &
Decl. & PosErr & $\Theta$ &
$NC$ & $BC$ & {PSF} &
PbNoSrc & EffExp & PbVar1 & 
PbVar2 & $ME$ & $F_X$\\
&&(deg)&(deg)&($\arcsec$)&($\arcmin$)&(cnts)&(cnts)&&&(ks)&&&(keV)&(ph/cm$^2$/s)\\
(1)&(2)&(3)&(4)&(5)&(6)&(7)&(8)&(9)&(10)&(11)&(12)&(13)&(14)&(15)\\
\cline{1-15}
&&&&&&&&&&&&&&\\ 
  1 & 213529.03+573412.3 & 323.871000 &  57.570087 & 0.5 & 11.1 &  $  130.9 \pm 12.9 $  & 10.1 & 0.90 &  0.0E+00 & 28.0 &  1.8E-01 &  1.8E-01 & 2.3 &  2.6E-05\\
  2 & 213532.88+572846.9 & 323.887026 &  57.479703 & 0.3 & 10.8 &  $  271.5 \pm 17.9 $  & 11.5 & 0.91 &  0.0E+00 & 29.1 &  3.2E-01 &  3.2E-01 & 1.5 &  4.9E-05\\
  3 & 213537.29+572842.0 & 323.905409 &  57.478345 & 1.2 & 10.2 &  $   16.3 \pm  6.3 $  & 10.7 & 0.91 &  6.7E-07 & 29.1 &  6.9E-01 &  6.9E-01 & 1.4 &  2.9E-06\\
  4 & 213538.86+573032.7 & 323.911957 &  57.509103 & 0.9 &  9.8 &  $   27.7 \pm  7.2 $  &  9.3 & 0.90 &  1.3E-12 & 28.9 &  8.9E-01 &  8.9E-01 & 2.0 &  5.0E-06\\
  5 & 213539.72+573127.5 & 323.915509 &  57.524320 & 1.1 &  9.1 &  $   13.7 \pm  5.6 $  &  6.3 & 0.89 &  6.2E-07 & 28.8 &  6.2E-01 &  6.2E-01 & 3.8 &  2.5E-06\\
  6 & 213541.87+573252.8 & 323.924497 &  57.548001 & 1.2 &  9.1 &  $    9.3 \pm  5.0 $  &  5.7 & 0.89 &  1.6E-04 & 28.9 &  9.9E-01 &  9.9E-01 & 2.9 &  1.7E-06\\
  7 & 213542.94+573334.8 & 323.928939 &  57.559685 & 0.9 &  9.1 &  $   19.1 \pm  6.2 $  &  6.9 & 0.90 &  4.9E-12 & 28.9 &  6.0E-01 &  6.0E-01 & 1.1 &  3.4E-06\\
  8 & 213543.96+572955.9 & 323.933171 &  57.498871 & 0.8 &  8.9 &  $   21.4 \pm  6.8 $  & 10.6 & 0.90 &  1.4E-08 & 48.4 &  2.0E-01 &  2.4E-01 & 1.7 &  2.2E-06\\
  9 & 213545.27+573326.5 & 323.938631 &  57.557375 & 1.0 &  8.8 &  $   11.9 \pm  5.4 $  &  6.1 & 0.90 &  4.2E-07 & 27.7 &  7.7E-01 &  7.7E-01 & 1.4 &  2.2E-06\\
 10 & 213550.45+573547.5 & 323.960213 &  57.596543 & 0.4 &  9.2 &  $  110.5 \pm 11.8 $  &  4.5 & 0.90 &  0.0E+00 & 20.7 &  2.3E-01 &  2.3E-01 & 1.5 &  2.7E-05\\
&&&&&&&&&&&&&&\\
\cline{1-15} 
\end{tabular}
\end{minipage}
\end{table*}

\begin{table*}\tiny
\centering
 \begin{minipage}{180mm}
 \caption{Optical and IR Photometry of X-ray Sources. This table is available in its entirety in the electronic edition of the journal. A portion is shown here for guidance regarding its form and content. Column 1: X-ray source number.  Columns 2-5: Optical $V_{FLWO}$, $I_{C,FLOW}$, $V_{LAICA}$, and $I_{J,LAICA}$  magnitudes. Columns 6-9: $2MASS$ $JHK_s$ magnitudes, and $2MASS$ photometry quality and confusion-contamination flags. Columns 10-14: IRAC magnitudes, and a digital flag giving photometric apertures and level of source contamination from nearby sources and nebular IR emission: ``3'' and ``4'' --- 3-pixel (2.6$\arcsec$) and 4-pixel (3.46$\arcsec$) apertures with contaminating flux from a neighboring source of no more than $5$\%; ``2'' and ``2c'' --- 2-pixel (1.73$\arcsec$) apertures with possible contaminating flux of $<10$\% and $>10$\%, respectively.}
 \label{tbl_oir_photometry}
 \begin{tabular}{@{\vline}|c@{ \vline}|c@{ \vline}c@{ \vline}c@{ \vline}c@{ \vline}c@{ \vline}c@{ \vline}c@{ \vline}c@{ \vline}c@{ \vline}c@{ \vline}c@{ \vline}c@{ \vline}c@{ \vline}}
\cline{1-14}
&&&&&&&&&&&&&\\
No. & $V_{FLWO}$ & $I_{C,FLWO}$ &
$V_{LAICA}$ & $I_{J,LAICA}$ & $J$ &
$H$ & $K_s$ & F$_2$ &
[3.6] & [4.5] &
[5.8] & [8.0] & F$_3$\\
&(mag)&(mag)&(mag)&(mag)&(mag)&(mag)&(mag)&&(mag)&(mag)&(mag)&(mag)&\\
(1)&(2)&(3)&(4)&(5)&(6)&(7)&(8)&(9)&(10)&(11)&(12)&(13)&(14)\\
&&&&&&&&&&&&&\\
\cline{1-14}
&&&&&&&&&&&&&\\
 1 & ... & ... & ... & ... & ... & ... & ... & ... & ... & ... & ... & ... & ...\\
  2 & ... & ... & ... & ... &  $  12.64 \pm   0.03 $  &  $  11.82 \pm   0.03 $  &  $  11.56 \pm   0.02 $  & AAA000 & ... & ... & ... & ... & ...\\
  3 & ... & ... & ... & ... &  $  13.65 \pm   0.03 $  &  $  12.95 \pm   0.03 $  &  $  12.71 \pm   0.03 $  & AAA000 & ... & ... & ... & ... & ...\\
  4 & ... & ... & ... & ... & ... & ... & ... & ... & ... & ... & ... & ... & ...\\
  5 & ... & ... & ... & ... & ... & ... & ... & ... &  $  16.20 \pm   0.06 $  &  $  15.64 \pm   0.05 $  & ... & ... & 2\\
  6 &  $13.44 \pm  0.00 $  &  $12.37 \pm  0.00 $  &  $13.57 \pm  0.00 $  &  $12.01 \pm  0.00 $  &  $  11.49 \pm   0.03 $  &  $  11.18 \pm   0.03 $  &  $  11.07 \pm   0.03 $  & AAA000 &  $  10.92 \pm   0.01 $  &  $  10.90 \pm   0.02 $  &  $  10.92 \pm   0.04 $  &  $  10.75 \pm   0.03 $  & 3\\
  7 & ... & ... &  $19.48 \pm  0.01 $  &  $15.51 \pm  0.00 $  &  $  14.30 \pm   0.04 $  &  $  13.44 \pm   0.03 $  &  $  13.07 \pm   0.04 $  & AAA000 &  $  12.53 \pm   0.00 $  &  $  12.28 \pm   0.00 $  &  $  12.13 \pm   0.01 $  &  $  11.54 \pm   0.04 $  & 4\\
  8 & ... & ... &  $19.44 \pm  0.01 $  &  $16.48 \pm  0.00 $  &  $  14.22 \pm   0.04 $  &  $  13.38 \pm   0.04 $  &  $  13.17 \pm   0.04 $  & AAA000 &  $  12.86 \pm   0.05 $  &  $  12.82 \pm   0.06 $  &  $  12.65 \pm   0.05 $  & ... & 2\\
  9 &  $19.50 \pm  0.08 $  &  $16.41 \pm  0.01 $  &  $19.36 \pm  0.01 $  &  $15.71 \pm  0.00 $  &  $  14.61 \pm   0.04 $  &  $  13.87 \pm   0.04 $  &  $  13.66 \pm   0.04 $  & AAA000 &  $  13.34 \pm   0.01 $  &  $  13.24 \pm   0.01 $  &  $  13.26 \pm   0.07 $  & ... & 4\\
 10 & ... & ... & ... & ... &  $  12.86 \pm   0.03 $  &  $  12.28 \pm   0.04 $  &  $  12.11 \pm   0.03 $  & AAAcc0 & ... & ... & ... & ... & ...\\
&&&&&&&&&&&&&\\
\cline{1-14}
\end{tabular}
\end{minipage}
\end{table*}

As X-ray emission from PMS stars is based on enhanced solar-type magnetic reconnection events rather than disk or accretion processes, X-ray selection delivers rich and clean samples of diskless stars missed by H$\alpha$ and IR selection \citep{Feigelson10}. Using archived X-ray {\it Chandra} grating data, \citet{Mercer09} identifed 22 new PMS stars around HD 206267, more than doubling the number of previously reported young stars in the central $10\arcmin \times 8\arcmin$ area of Tr~37. We present here the identification of $>130$ previously unknown members of the Tr 37/IC 1396A stellar populations, using new X-ray and optical observations with archived {\it Spitzer} data. The {\it Chandra} and the auxiliary {\it Spitzer} and optical FLWO/LAICA data are described in \S \ref{data_reduction_section}. Identification of X-ray sources with optical-IR counterparts and membership classification are considered in \S \ref{ir_optical_counterparts_section}. Classification of diskbearing and diskless X-ray emitting YSOs is provided in \S \ref{disk_classes_subsection}, and the discovery of a new population of non-{\it Chandra} IR-excess members is given in \S \ref{non_chandra_members_subsection}. We then estimate the global properties of the Tr~37/IC~1396A stellar populations including age distribution (\S \ref{age_analysis_section}), spatial structure (\S \ref{spatial_structure_section}), X-ray luminosity functions (XLFs) and initial mass functions (IMFs; \S \ref{total_populations_section}). We end with a discussion of the implications for understanding and quantifying triggered star formation in the region (\S \ref{implications_for_tsf_section}).

\begin{table*}\tiny
\centering
 \begin{minipage}{180mm}
 \caption{Derived Properties, Stellar Identifications, Membership, and Classification. This table is available in its entirety in the electronic edition of the journal. A portion is shown here for guidance regarding its form and content. Column 1: X-ray source number. Columns 2-3: X-ray net counts and median energy for the full $(0.5-8.0)$~keV band. Columns 4-5: X-ray column density, intrinsic luminosity, and their errors (summed in quadrature statistical and systematic errors). Luminosities are derived assuming a distance of $870$~pc and are given for the full $(0.5-8.0)$~keV band. Columns 6-7: Age estimates derived from the optical $V$ $vs.$ $V-I_C$ and $V$ $vs.$ $V-I_J$  color magnitude diagrams using PMS models of \citet{Siess00} and assuming a constant source extinction for optical PMS candidates of $A_V = 1.56$~mag (SA05, B11). Column 8: Apparent SED slope from IRAC photometry with $1\sigma$ error. Column 9: Number of IRAC bands, from which the SED slope was derived. Column 10: $2MASS$ identifier. Column 11: Source identifier from our analysis of the IRAC data; X-ray sources outside the field of view of the IRAC data analyzed here are labeled as ``OutOfIracFOV''. Columns 12-14: Stellar counterparts from SA05, B11, and MC09. Column 15: This positional flag indicates if an X-ray source lies projected against the globule (``1'') or not (``0''). Column 16: Indicates membership and YSO class: ``EXG'' --- possible extragalactic contaminant; ``FRG'' --- possible foreground contaminant or YSO; ``DSK'' and ``NOD'' --- disky and diskless YSOs, respectively; ``UNC1'' and ``UNC2'' --- objects of the uncertain class.  ``UNC1'' are sources with no IR counterparts; these are most likely extragalactic objects but some can be background stars. ``UNC2'' have weak registered and unregistered IR counterparts or lie close to bright IR objects; these could be YSOs or field stars or extragalactic objects.}
 \label{tbl_derived_props}
 \begin{tabular}{@{\vline}|c@{ \vline}|c@{ \vline}c@{ \vline}c@{ \vline}c@{ \vline}c@{ \vline}c@{ \vline}c@{ \vline}c@{ \vline}c@{ \vline}c@{ \vline}c@{ \vline}c@{ \vline}c@{ \vline}c@{ \vline}c@{ \vline}}
\cline{1-16}
&&&&&&&&&&&&&&&\\

No. & $NC$ & $ME$ &
$\log(N_H)$ & $\log(L_X)$ & $t_{FLWO}$ & $t_{LAICA}$ &
$\alpha_0$ & $Np$ &
2MASS & IRAC &
SA05 & B11 &
MC09 & Region & Class\\

&(cnt)&(keV)&(cm$^{-2}$)&(erg/s)&(Myr)&(Myr)&&&&&&&&&\\
(1)&(2)&(3)&(4)&(5)&(6)&(7)&(8)&(9)&(10)&(11)&(12)&(13)&(14)&(15)&(16)\\
&&&&&&&&&&&&&&&\\
\cline{1-16}
&&&&&&&&&&&&&&&\\
164 &    7.8 & 1.8 & ... & ... & ... & ... &  $  -1.36 \pm   0.10 $  & 4 & 21364762+5729540 & J213647.61+572954.1 & ... &  30 & IC1396A-61 & 1 & DSK\\
165 &    3.4 & 2.7 & ... & ... & ... & ... & ... & ... & ... & ... & ... & ... & ... & 0 & UNC1\\
166 &   18.9 & 4.1 & $ 23.04 \pm 0.21 $ & $ 30.97 \pm 0.24 $ & ... & ... &  $  -1.15 \pm   0.20 $  & 4 & 21364788+5731306 & J213647.87+573130.6 & ... & ... & IC1396A-28 & 1 & DSK\\
167 &   17.9 & 1.3 & $ 21.60 \pm 0.50 $ & $ 29.71 \pm 0.23 $ & ... & ... &  $  -2.43 \pm   0.07 $  & 3 & 21364793+5723062 & J213647.94+572306.5 & ... & ... & ... & 0 & NOD\\
168 &   12.5 & 1.3 & ... & ... &   4.7 &   2.8 &  $  -2.54 \pm   0.07 $  & 2 & 21364819+5734020 & J213648.19+573401.9 & ... & ... & ... & 0 & NOD\\
169 &   59.0 & 1.1 & $ 20.15 \pm 0.77 $ & $ 29.92 \pm 0.09 $ & ... & ... &  $  -2.83 \pm   0.01 $  & 4 & 21364825+5739185 & J213648.25+573918.4 & ... & ... & ... & 0 & FRG\\
170 &   11.7 & 3.1 & $ 22.58 \pm 0.26 $ & $ 30.61 \pm 0.30 $ & ... & ... & ... & ... & ... & OutOfIracFOV & ... & ... & ... & 0 & UNC2\\
171 &   12.8 & 1.3 & $ 21.60 \pm 0.69 $ & $ 29.48 \pm 0.25 $ & ... & ... &  $  -2.25 \pm   0.11 $  & 3 & 21364866+5730262 & J213648.65+573026.2 & ... & ... & ... & 1 & NOD\\
172 &    6.9 & 1.0 & ... & ... & ... & ... & ... & ... & ... & ... & ... & ... & ... & 0 & UNC1\\
173 &   15.9 & 2.6 & $ 22.41 \pm 0.17 $ & $ 30.48 \pm 0.25 $ &   0.6 &   0.3 &  $  -0.94 \pm   0.01 $  & 4 & 21364941+5731220 & J213649.42+573122.2 & 14-141 & ... & IC1396A-29 & 1 & DSK\\
174 &   96.1 & 1.5 & $ 21.78 \pm 0.21 $ & $ 30.56 \pm 0.16 $ &   3.5 &   3.8 &  $  -1.76 \pm   0.02 $  & 4 & 21364964+5722270 & J213649.65+572227.1 & ... & ... & ... & 0 & DSK\\
&&&&&&&&&&&&&&&\\
\cline{1-16}
\end{tabular}
\end{minipage}
\end{table*}

\section{{\it CHANDRA}-ACIS, {\it SPITZER}-IRAC, AND OPTICAL OBSERVATIONS}\label{data_reduction_section}

\subsection{X-ray Data}\label{xray_data_subsection}

The current project consists of two {\it Chandra}-ACIS \citep{Weisskopf02,Garmire03} X-ray observations of IC 1396A, a Guaranteed Time observation (ObsID No. 11807 obtained on 2010 March 31; PI Garmire) and a Guest Observer observation (ObsID No. 10990 obtained on 2010 June 9; PI Getman). Both observations were pointed at the head of the globule but had different roll angles\footnote{The aim points of the observations were $21^{\rm{h}}36^{\rm{m}}51\fs5$, $+57\degr30\arcmin51\farcs3$ (J2000) and $21^{\rm{h}}36^{\rm{m}}47\fs6$, $+57\degr30\arcmin43\farcs3$ (J2000) for ObsIDs 11807 and 10990, respectively. The roll angles were $52\fdg4$ and $121\fdg2$ for ObsIDs 11807 and 10990, respectively.}. For each observation, we consider here only results arising from the imaging array (ACIS-I) of four abutted $1024 \times 1024$ pixel front-side illuminated charge-coupled devices (CCDs) covering about $17\arcmin \times 17 \arcmin$ on the sky. The S2 and S3 detectors in the spectroscopic array (ACIS-S) were also operational, but as the telescope point spread function (PSF) is considerably degraded far off-axis, the S2-S3 data are omitted from the analysis. The overlapping field of the two observations achieved 59~ks of exposure with no data losses or background flaring due to solar activity.

Data reduction follows procedures similar to those described in detail by \citet{Broos10,Broos11}. Level 1 event data were calibrated and cleaned using mostly standard methods and tools (CIAO version 4.3; CALDB version 4.4.0).  Event positions were adjusted to better align with the Naval Observatory Merged Astrometric Dataset catalog \citep[NOMAD;][]{Zacharias04}. The effective point-spread function was slightly improved by removing a pixel randomization that was historically added to ACIS event positions, and by applying a sub-pixel position estimation algorithm \citep{Li04}. A custom bad pixel table was used to slightly increase point source sensitivity. Instrumental background events were identified and removed using an aggressive algorithm when searching for sources and using an algorithm with few false positives when extracting sources \citep[][their Figure 1]{Broos10}.

Detection of point sources is performed using both the wavelet transform method \citep{Freeman02} and the Lucy-Richardson image reconstruction algorithm \citep[][their \S 4.2]{Broos10}; the latter is better suited for resolving closely spaced sources. Updated position estimates, photometry, and spectra were obtained for candidate sources using the {\em ACIS Extract} (AE) software package\footnote{The {\em ACIS Extract} software package and User's Guide are available at http://www.astro.psu.edu/xray/acis/acis\_analysis.html .} \citep{Broos10,Broos12}.

Some of the inferred properties, such as X-ray net counts ($NC$), background counts ($BC$), median energy ($ME$), and apparent flux ($F_X$) for the full ($0.5-8$)~keV energy band, as well as $p$-value for no-source hypothesis \citep[$P_bNoSrc$;][their \S 4.3 and Appendix B]{Broos10}, are provided in Table \ref{tbl_xray_photometry}. Similar to the {\it Chandra} catalog of X-ray sources in the Carina Nebula \citep{Broos11}, our list of candidate sources in IC 1396A is trimmed to omit sources with fewer than 3 total source counts ($NC+BC <3$) and probability for being a background fluctuation greater than 1\% ($P_bNoSrc > 0.01$).

Our final catalog comprises 415 X-ray sources (Table \ref{tbl_xray_photometry}). We expect that roughly half of these sources are extragalactic with extremely optically-faint counterparts (\S \ref{membership_subsection}), and the rest are young stars associated with the Tr~37 and IC~1396A star forming regions.

\subsection{IRAC Data}\label{irac_data_reduction_subsection}

To obtain MIR photometry for X-ray objects and to identify and measure MIR photometry for additional non-{\it Chandra} disky stars that were missed in previous studies of the region, we have reduced the archived {\it Spitzer}-IRAC \citep{Fazio04} data from a Guaranteed Time observation of the program \#37 (Fazio, PI). The same observation has been presented in SA06 and \citet{Sicilia-Aguilar11}[hereafter SA11]. The SA06 data had been reduced using old calibration and flat fielding information, and the photometry for sources with highly non-uniform background was additionally affected by the choice of large source extraction apertures. Given the improvement of the pipeline and calibration since then, SA11 have re-reduced the data and redone the photometry for a selected sample of YSOs. Since both SA06 and SA11 catalogs lack photometry measurements for the bulk of the X-ray sources, reanalysis of the data is in order.

\begin{figure*}
\centering
\includegraphics[angle=0.,width=170mm]{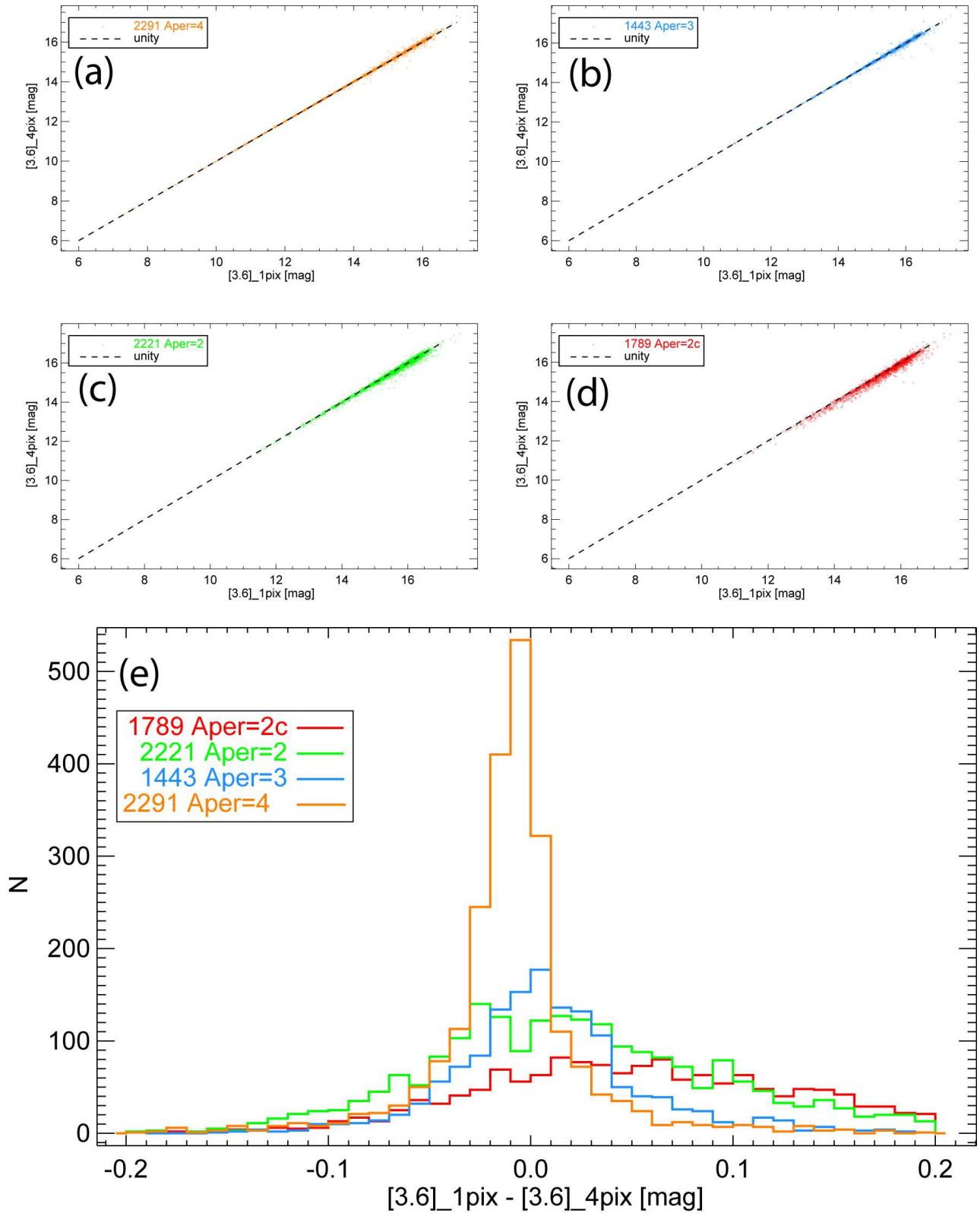}
\caption{Entire IRAC source sample. [3.6] band. The photometry from 4 pixel wide background annuli versus the photometry from 1 pixel wide background annuli. The top 4 panels show the photometric measurements for the source samples with different source extraction apertures: 4-pix (orange), 3-pix (blue), 2-pix (green), and 2-pix with possible contaminating flux from a neighboring source of $> 10$\% (red). The bottom panel is a histogram representation of the photometric differences for all 4 source samples with different source extraction apertures. The median, mean, and standard deviation of the magnitude differenceas are: 0.069, 0.094, and 0.133~mag for the ``2c'' extractions; 0.02, 0.034, and 0.099~mag for the 2-pix extractions; 0.007, 0.011, and 0.077~mag for the 3-pix extractions; and -0.008, -0.008, and 0.06~mag for the 4-pix extractions.  \label{fig_irac_2new}}
\end{figure*} 

The observation was obtained on 2003 December 20  with the IRAC detector in all four IRAC channels, 3.6, 4.5, 5.8, and 8.0~$\mu$m. Two adjacent fields subtending $\sim 37\arcmin \times 42\arcmin$ in channel pairs 3.6/5.8~$\mu$m and 4.5/8.0~$\mu$m were centered on Tr~37. These fields were imaged with an $8 \times 9$ mosaic of adjacent positions separated by $280\arcsec$; individual exposures were taken at 5 dithered positions for each of the 72 cells of the mosaic. The images were taken in the high-dynamic-range mode with 30 and 1.2 second frame time exposures. To reduce unnecessary data processing we have analyzed only a portion of the original data that encompasses the {\it Chandra}-ACIS field with the coverage of $\sim 19\arcmin \times 19\arcmin$ area in all four channels centered on the Rim A of the IC 1396A globule. This covers $93$\% of the ACIS field omitting its north-western and south-western edges.

Basic Calibrated Data (BCD) products from the {\it Spitzer} Science Center's IRAC pipeline version S18.18.0 were automatically treated with the WCSmosaic IDL package developed by R. Gutermuth from the IRAC instrumental team. Starting with BCD data products, the package mosaics individual exposures while treating bright source artifacts, cosmic ray rejection, distortion correction, subpixel offsetting, and background matching \citep{Gutermuth08}. We selected a plate scale of 0.86$\arcsec$ for the reduced IRAC mosaics, which is the native scale divided by $\sqrt{2}$. 

Aperture photometry of IRAC sources was obtained using the IRAF task PHOT. Circular apertures with radius of 2, 3, and 4 pixels ($1.72\arcsec$, $2.58\arcsec$, $3.44\arcsec$) are assigned, with larger apertures favored for uncrowded sources and smaller apertures favored for crowded sources. Using the IRAC point-spread function (PSF) for the [3.6] band resampled to a plate scale of 0.86$\arcsec$, pairs of sources are simulated with wide ranges of source separations, orientations, and flux ratios to derive the flux contribution from a nearby source within 2, 3, and 4 pixel apertures as a function of separation angle and flux ratio (Kuhn et al. in preparation). For the real sources, their photometry is derived using 2, 3, and 4 pixel source apertures, and we report photometry from the largest aperture for which the expected contamination from a nearby source is less than 5\%. A source aperture radius of 4 pixels was used for most of the X-ray sources and non-{\it Chandra} selected sources of interest (\S \ref{non_chandra_members_subsection}). The cases for which the expected contamination is $>10$\% in the smallest 2 pixel aperture are flagged in Table \ref{tbl_oir_photometry} as ``2c''.

The background in the region is highly non-uniform due to the contribution by the PSF cores and wings of multiple neighboring sources often superimposed on top of the spatially variable diffuse emission. For all IRAC sources the aperture photometry was performed for two different sizes of sky annuli, 4 pixel and 1 pixel wide sky annuli adjoining source extraction regions. The former, closer in size to the ``standard'' sky annuli recommended in the IRAC Instrument Handbook, collects higher numbers of sky pixels offering smaller statistical uncertainties but with the danger of increased systematic error due to sampling wrong background. The latter is chosen in accordance with the strategy of e.g., \citet{Lada06, Luhman08, Getman09} and K. Luhman(private communication) who attempt to sample the actual sky present within the source extraction region. For background selection, the optimal balance between proximity and number of background pixels is not obvious.

We adopt zero point magnitudes ($ZP$) of 19.670, 18.921, 16.855, and 17.394 in the 3.6, 4.5, 5.8, and 8.0~$\mu$m bands, where $M = -2.5 \log (DN/sec) + ZP$ \citep{Reach05}. Since our plate scale and the extraction apertures have custom sizes, the aperture correction values are calculated here using the current data as desribed by \citet{Luhman10}. The calculations are based on the comparison of the photometric measurements for the relatively isolated, bright, and unsaturated sources obtained for the source and sky regions of different sizes including large 14 pixel ($\sim 10$ native pixel) extraction regions that are similar in size to those used in the calibrations of \citet{Reach05}. The total aperture corrections applied to our measurements in the case of 4-pix background annulus are: 0.213, 0.184, 0.218, 0.416 mag in the case of 4-pixel source aperture; 0.441, 0.367, 0.698, 0.949 mag in the case of 3-pixel source aperture; and 0.829, 0.955, 1.319, 1.105 mag in the case of 2-pixel source aperture in the 3.6, 4.5, 5.8, and 8.0~$\mu$m bands, respectively. The total aperture corrections applied to our measurements in the case of 1-pix background annulus are: 0.174, 0.171, 0.161, 0.180 mag in the case of 4-pixel aperture; 0.387, 0.304, 0.496, 0.653 mag in the case of 3-pixel aperture; and 0.637, 0.714, 0.976, 0.868 mag in the case of 2-pixel aperture in the 3.6, 4.5, 5.8, and 8.0~$\mu$m bands, respectively. The reported photometric errors include statistical errors in the source and background emission and a 2\% uncertainty in the calibration of IRAC \citep{Reach05}. For further analyses we retain IRAC sources with signal-to-noise $>3$ in both the [3.6] and [4.5] bands, and the photometry is reported if the signal-to-noise $>3$.

For all extracted IRAC sources we compare their photometric measurements performed with 1-pix and 4-pix background annuli. Figure \ref{fig_irac_2new} exemplifies such a comparison for the [3.6] band. For the sources with source apertures 4-pix (orange), 3-pix (blue), and 2-pix (green) the distributions of the magnitude differences typically have small biases of $<0.02-0.05$~mag and dispersions of $<0.04-0.2$~mag increasing towards the fainter magnitudes and longer wavelengths. Both the effect of the finite number of background pixels and the possible physical mistakes the two methods are making could be the cause of the magnitude biases and spreads.  In the case of the most crowded sources, with the apertures ``2c'' (red), the bias and dispersion are higher, $<0.02-0.08$~mag and $<0.1-0.4$~mag, respectively. The nature of this bias is unclear.  For example, if a neighbor lies within the 1~pixel aperture, then the larger 4~pixel aperture would dilute the light from this neighbor, shifting the photometry in the observed direction. For the X-ray sub-sample of ``2c'' sources the bias appears only in the range of $[3.6]>16$ mag characteristic of objects unrelated to this relatively close star forming region (\S \ref{membership_subsection}). Performing the disk classification based on spectral energy distributions (SEDs) as described in \S \ref{disk_classes_subsection}, we find that the choice of the background aperture appears to have a tiny effect on SED shapes and no discernible effect on disk classification for all X-ray YSOs. Since there is no evident preference for either background annulus size, we publish the photometry from the method with the 1 pixel wide background annuli, which produces conservatively large statistical errors. Derived IRAC photometric measurements for 242 X-ray sources are given in Table \ref{tbl_oir_photometry}.

\begin{figure*}
\centering
\includegraphics[angle=0.,width=170mm]{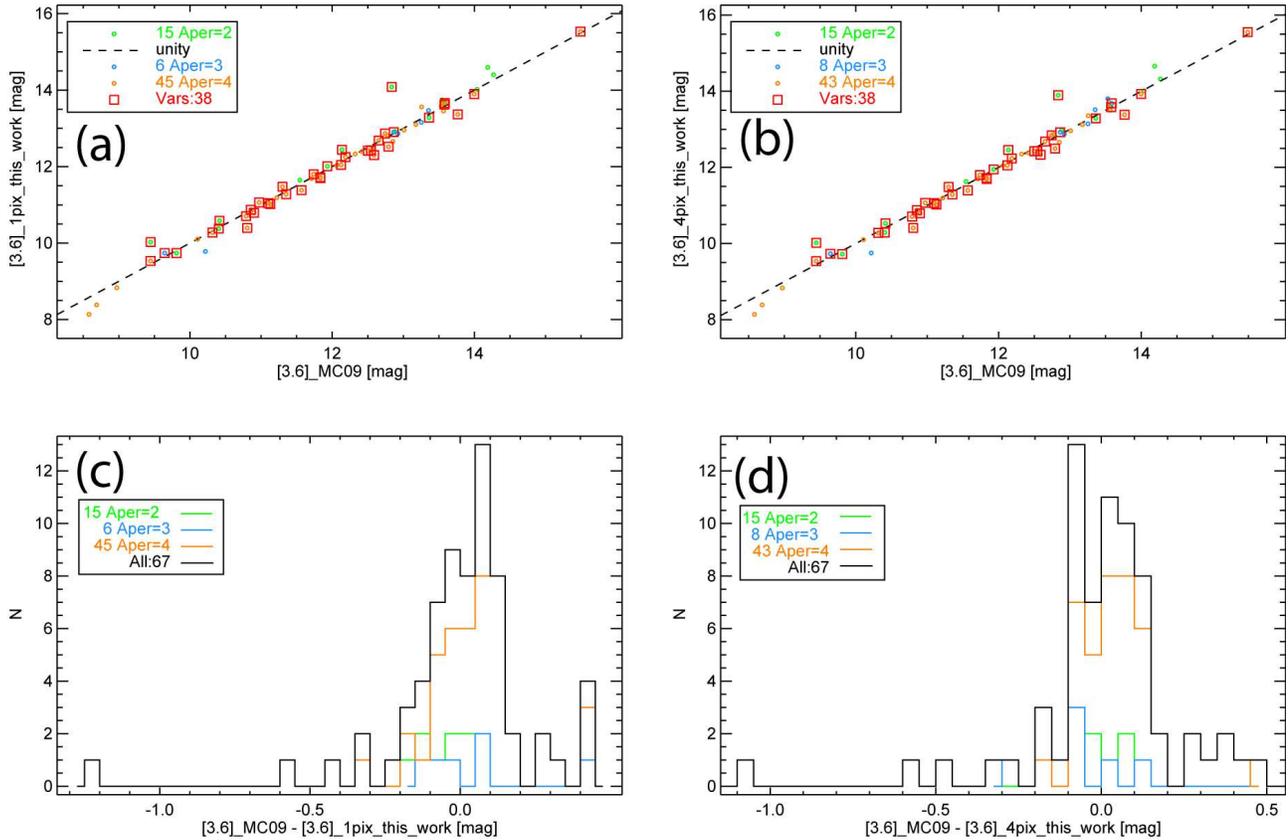}
\caption{Photometry comparison with the time-averaged measurements of MC09. [3.6]-band. Left (right) panels present the photometric measurements for our 1-pix (4-pix) background methods versus MC09 photometry. The top and bottom panels give a scatter-plot and a histogram versions, respectively. The source samples are color-coded according to our choice of different source extraction apertures: 4-pix (orange), 3-pix (blue), and 2-pix (green). The MIR-variable sources from MC09 are additionally marked by the red squares. On panel (c) the median, mean, and standard deviation of the magnitude differenceas are: -0.102, -0.207, and 0.341~mag for the 2-pix extractions; 0.062, 0.057, and 0.203~mag for the 3-pix extractions; 0.056, 0.064, and 0.153~mag for the 4-pix extractions; and 0.026, 0.003, and 0.237~mag for the entire sample. On panel (d) the median, mean, and standard deviation of the magnitude differenceas are: -0.082, -0.176, and 0.316~mag for the 2-pix extractions; -0.058, -0.007, and 0.224~mag for the 3-pix extractions; 0.056, 0.07, and 0.141~mag for the 4-pix extractions; and 0.024, 0.006, and 0.222~mag for the entire sample. \label{fig_irac_3new}}
\end{figure*} 

For the objects in common with MC09 we compare our photometric measurements with the time-averaged measurements of MC09 that are based on new deeper {\it Spitzer}-IRAC exposures. Figure \ref{fig_irac_3new} exemplifies such a comparison for the [3.6] band.  The source sample here is relatively bright ($[3.6]<14$ mag), so the magnitude differences are not affected by the choice of our background apertures. The magnitude differences have small biases of $0.02-0.03$ mag and dispersions of $0.22-0.25$ mag. About 60\% of the sources considered here are MIR variable. Since the sources are relatively bright, variability rather than photometric uncertainty is likely the major cause of the magnitude spread. For the 28 X-ray YSOs with MC09 counterparts, our SED-based disk classification agrees with that of MC09 for all but one source (\S \ref{disk_classes_subsection}).

{\it Spitzer}-MIPS data are also available (PID 58; PI Rieke). The use of the MIPS data in the current work could potentially lead to revised disk classes for some of the sources --- from diskless classified by shorter wavelengths to Transition Disks (TDs), systems with inner disk holes. However, due to the reduced spatial resolution and sensitivity of the MIPS data such a classification would be biased towards brighter, less crowded objects located mainly outside the globule (the region with the brightest diffuse emission). For simplicity, in this work we prefer to omit the usage of the MIPS data and to classify any TD candidate as diskless (\S \ref{disk_classes_subsection}). This should not affect the goals of the paper, identification and characterization of triggered stellar populations.

\subsection{FLWO and Calar Alto Data}\label{laica_data_reduction_subsection}

The optical observations of the Tr~37 and IC 1396A and their data analyses are described in detail by \citet{Sicilia-Aguilar04, Sicilia-Aguilar05, Sicilia-Aguilar10}. 

The $UVR_CI_C$ observations were carried on with the 1.2~m telescope at the Fred Lawrence Whipple Observatory (FLWO), using the 4Shooter CCD array, between September 2000 and September 2002 \citep{Sicilia-Aguilar04, Sicilia-Aguilar05}. The 4Shooter is an array of four CCDs, covering a square of $25\arcmin$ on the side. Two 4Shooter fields were taken to cover $\sim 45\arcmin \times 25\arcmin$ area centered on HD~206267. The FLWO fields contain the whole ACIS field, except for a small gap in between the 4 CCDs of the 4Shooter. All but a few {\it Chandra} stars were observed in September 2000.

The $UVR_JI_J$ observations of Tr~37/IC 1396A were obtained in service mode during three nights in 2007 June 9-11, using the Wide-Field Camera LAICA mounted on the 3.5~m telescope in Calar Alto, Spain \citep{Sicilia-Aguilar10}. LAICA is a $2 \times 2$ mosaic of four CCDs, each covering a $15.3\arcmin \times 15.3\arcmin$ field of view with a large gap of $15.3\arcmin \times 15.3\arcmin$ in between. The project combines four LAICA pointings covering $\sim 45\arcmin \times 45\arcmin$ area around HD~206267, including nearly the entire ACIS field.

After applying standard procedures for bias and flat field corrections, aperture photometry, and zero point calibration \citep{Sicilia-Aguilar04, Sicilia-Aguilar10}, we obtain optical photometry for 181 X-ray sources. The resulted $VI_C$ and $VI_J$ photometric measurements are given in Table \ref{tbl_oir_photometry}.

\section{INFRARED AND OPTICAL COUNTERPARTS TO X-RAY SOURCES}\label{ir_optical_counterparts_section}

\subsection{Identifications}\label{identifications_subsection}

Positions of our X-ray sources were compared with source positions from the three optical catalogs (SA05/FLWO, B11, and LAICA), the NIR 2MASS catalog, and the two MIR {\it Spitzer} catalogs (MC09 and our custom catalog from \S \ref{irac_data_reduction_subsection}). SA05, B11, LAICA, and 2MASS catalogs cover the full ACIS field. Our custom MIR catalog overlaps $93$\% of the ACIS field omitting its north-western and south-western edges. MC09 catalog covers the inner region of the ACIS field near/inside the globule; this amounts to $\sim 20$\% of the ACIS field. MC09 catalog includes all diskbearing stars near/inside the globule reported by \citet{Reach04} and \citet{Sicilia-Aguilar06}.

\begin{figure*}
\centering
\includegraphics[angle=0.,width=130mm]{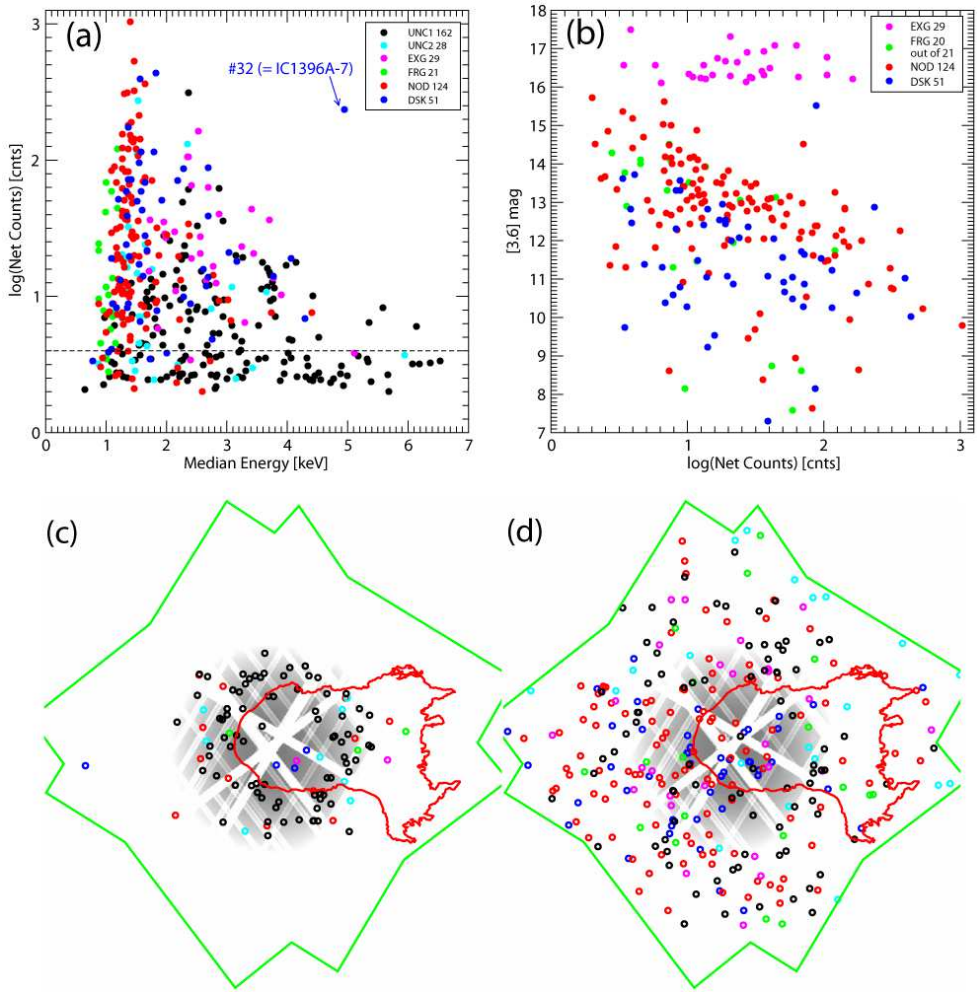}
\caption{Properties of different classes of X-ray sources. Source classes are color-coded: the uncertain class 1 (UNC1), which is largely comprised of extragalactic objects, is in black; the uncertain class 2 (UNC2), which is possibly a mixture of  contaminants and some low-mass YSOs, is in cyan; the EXG class, which is composed of IR-X-ray bright extragalactic objects, is in magenta; the FRG class that likely contains foreground stars and some YSOs is in green; and the NOD/DSK classes are composed of young dikless/disky members shown in red/blue. (a) X-ray net counts versus X-ray median energy. The dashed line marks the 4 net count level. One of the most reddest Class~I protostars found within the globule ({\it Chandra} source \#32 = IC1396A-7) is labeled. (b) $3.6$~$\mu$m-band magnitude versus X-ray net counts. (c)-(d) Spatial distribution of different classes of X-ray sources superposed on the  {\it Chandra} exposure map shown in inverted colors with logarithmic scales. The distributions are stratified by X-ray net counts, $NC<4$ counts (c) and $NC>4$ counts (d). The green polygon outlines the {\it Chandra} FOV. The red contour demarcates the globule, tracing $8.0$~$\mu$m emission from hot dust and PAHs. \label{fig_mem}}
\end{figure*}

An automated cross-correlation between the {\it Chandra} source positions and optical-IR source positions was made using a search radius of $2\arcsec$ within $\sim 6\arcmin$ of the ACIS field center, and a search radius of $3.5\arcsec$ in the outer regions of the ACIS field where X-ray source positions are more uncertain due to the deterioration of the {\it Chandra} telescope point spread function. This was followed by a careful visual examination of each source in both bands to remove dubious sources and associations.  

Optical FLWO/LAICA (\S \ref{laica_data_reduction_subsection}), NIR 2MASS, and MIR IRAC (\S \ref{irac_data_reduction_subsection}) photometry for X-ray sources is given in Table \ref{tbl_oir_photometry}. Individual counterparts from 2MASS, our custom IRAC (\S \ref{irac_data_reduction_subsection}), SA05, B11, and MC09 catalogs are given in Table \ref{tbl_derived_props}.

{\it Chandra} detects 27 out of 69 (39\%) of the MC09 infrared-excess objects and 45 out of 76 (59\%) of the B11/SA05 H$\alpha$ emitters and Li absorbers. Most of the optical YSOs undetected in X-rays are diskbearing stars. The X-ray detection efficiency of diskbearing stars is known to be lower than that of diskless stars \citep[][and Figure \ref{fig_JvsJH_acis} here]{Getman09,Stelzer11}. To compensate for this effect, in section \S \ref{non_chandra_members_subsection} we introduce a catalog of additional non-{\it Chandra} IR-excess selected disky stars within the ACIS field and beyond. This allows a recovery of 83\% of H$\alpha$ emitters and Li absorbers. Several optical stars located at the very edge of the ACIS field, where the ACIS sensitivity degrades, remain unrecovered.

\subsection{Pre-Main-Sequence Census}\label{membership_subsection}

X-ray surveys of star-forming regions suffer contamination by extragalactic sources, mainly quasars and other active galactic nuclei, which can be seen even through the Galactic plane as faint, absorbed X-ray sources. Additional contamination arises from foreground and background Galactic stars, mainly main-sequence stars and some types of giants \citep[][and references therein]{Getman11}. 

We perform detailed simulations for extragalactic and Galactic X-ray contamination populations expected in the direction of the IC 1396A ACIS field, $(l,b)=(99.1\degr, +3.9\degr)$.  The methodology for such simulations is detailed in \citet{Getman11}. The simulations take into consideration a variety of factors involving a Galactic population synthesis model \citep{Robin03}, stellar (AGN) X-ray luminosity (flux) functions, Chandra telescope response, source detection methodology, and possible spatial variations in the X-ray background and absorption through molecular clouds. For the IC 1396A cloud, we construct its extinction map based on dust redenning of 2MASS sources excluding possible foreground sources, all previously known YSOs, and 2MASS sources with X-ray counterparts \citep[for detailed methodology see Appendix B in][]{Schneider11}. The derived extinction varies from $A_V<2$~mag outside the globule to $3 \la A_V \la 12$~mag through the head of the main globule (behind Rim~A) and along the edge of SFO~36 cloudlet adjacent to Rim~Aa (Figure \ref{fig_intro}). Our simulations predict that $\sim 20-30$ foreground X-ray stars with X-ray median energies of $ME \la 1$~keV, $\sim 20-30$ background stars with $1<ME<2$~keV, and $\sim 110-150$ extragalactic objects with $ME>2$~keV are present in the ACIS field.

Based on presence/absence of IR counterparts, source spatial distributions, X-ray counts, X-ray median energies (surrogates for X-ray absorption), and the predicted number of contaminants, the X-ray objects detected in our {\it Chandra} observation can be separated into the following source classes (see Table \ref{tbl_derived_props} and Figure \ref{fig_mem}):
 
\begin{enumerate}

\item[A.] Bright extragalactic objects (``EXG'').

\item[B.] Uncertain class 1 (labeled as ``UNC1'') mostly composed of faint extragalactic objects.

\item[C.] Uncertain class 2 (``UNC2''), which is possibly a mixture of contaminants and some low-mass YSOs.

\item[D.] A mixture of foreground stars and possibly some YSOs (``FRG'').

\item[E.] Diskless and disky young members of the region (``NOD'' and ``DSK'').

\end{enumerate}

The 29 hard X-ray sources ($ME \ga 2$~keV; magenta in Figure \ref{fig_mem} (a)) from the class ``EXG'' lack 2MASS but have dim MIR counterparts ($[3.6] > 16$~mag, $[4.5]>14.5$~mag). These are most likely bright extragalactic objects, because they are clear outliers from the stellar locus on the $[3.6]$ versus X-ray net count diagram (Figure \ref{fig_mem} (b)), and their $[3.6]$ and $[4.5]$ magnitudes and X-ray fluxes are fully consistent with those of the {\it Chandra}-{\it Spitzer} AGN sample from the Serendipitous Extragalactic X-ray Source Identification (SEXSI) program \citep{Eckart10}.

\begin{figure*}
\vspace*{0.2in}
\centering
\includegraphics[angle=0.,width=120mm]{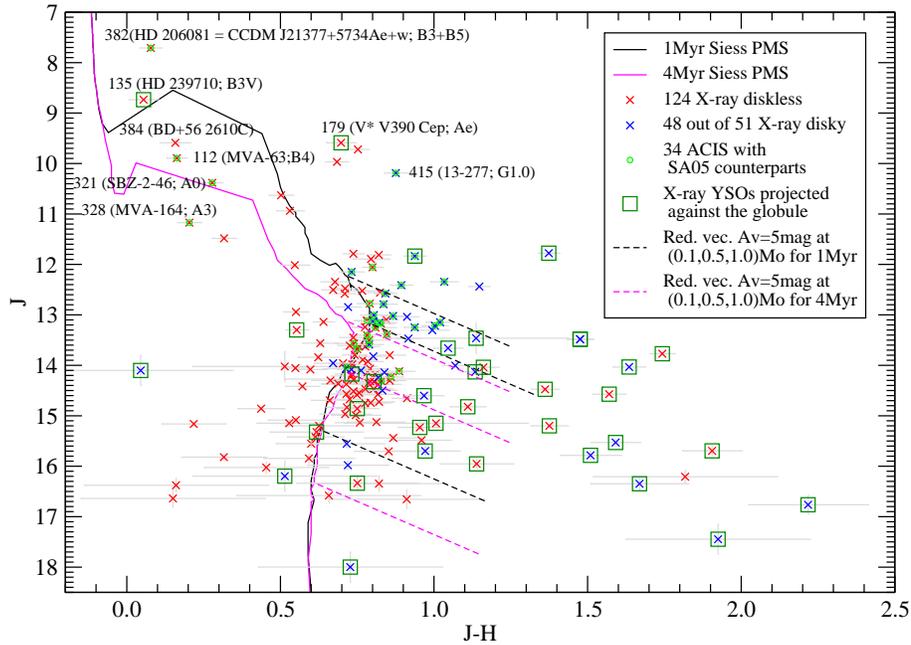}
\caption{NIR color-magnitude diagram for all $Chandra$ YSOs with available NIR photometry. Diskless and disky X-ray stars are marked as red and blue $\times$, respectively. Sources projected against the globule are further outlined by green $\sq$. X-ray sources with known optical counterparts from SA05 are additionally marked by green $\circ$. 1 and 4~Myr PMS isochrones from \citet{Siess00} at $870$~pc distance are shown as solid black and magenta lines, respectively. Reddening vectors of $A_V = 5$~mag for various masses are indicated by dashed lines.  Some previously known intermediate and high-mass stars are labeled. \label{fig_JvsJH_acis}}
\end{figure*}

\begin{figure*}
\vspace*{0.5in}
\centering
\includegraphics[angle=0.,width=120mm]{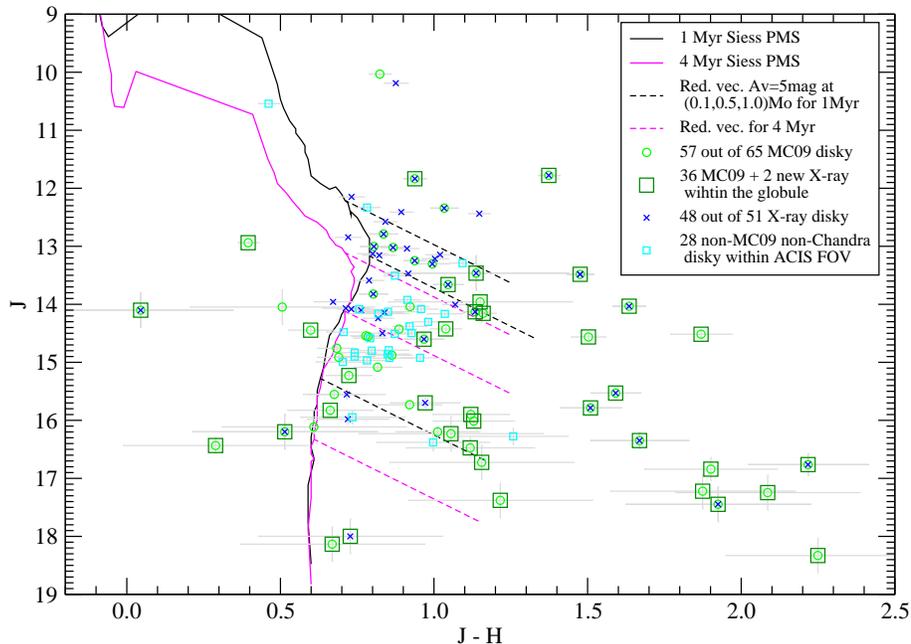}
\caption{NIR color-magnitude diagram for all $Chandra$+$Spitzer$ disky stars with available NIR photometry identified within the ACIS field. X-ray stars are marked by blue $\times$. Disky stars from MC09 are marked by green $\circ$. Additional non-{\it Chandra} and non-MC09 disky stars (from \S \ref{non_chandra_members_subsection}) within the ACIS field are marked as cyan $\sq$. Stars projected against the globule are further outlined by green $\sq$. 1 and 4~Myr PMS isochrones from \citet{Siess00} at $870$~pc distance are shown as solid black and magenta lines, respectively. Reddening vectors of $A_V = 5$~mag for various masses are indicated by dashed lines. \label{fig_JvsJH_disky}}
\end{figure*}

The class ``UNC1'' is composed of 162 X-ray sources. We believe that most of these sources are extragalactic objects because they lack IR counterparts, are generally X-ray faint and hard ($NC<4-10$ counts and $ME \ga 2$~keV; Figure \ref{fig_mem} (a)), and their number is comparable to the number of simulated AGNs. Their spatial distribution is also similar to that of the simulated extragalactic objects: the weakest X-ray sources are concentrated towards the central part of the ACIS field while brighter sources are uniformly distributed across the field (black points in Figure \ref{fig_mem} (c)-(d)). The greater density of very weak X-ray sources on-axis than off-axis is due to variation in detection completeness with off-axis angle \citep{Broos11}. The detected X-ray YSOs are much less prone to such an effect because only a handful of them are very weak sources (red and blue points in Figure \ref{fig_mem}).

The 28 objects of the class ``UNC2'' do not have cataloged IR counterparts, but our visual inspection of the 2MASS and {\it Spitzer}-IRAC images suggests that some might have very dim uncatalogued counterparts and some lie near bright IR objects. We suggest that this class is a mixture of heterogeneous contaminants and low-mass YSOs.

The remaining X-ray sources with IR counterparts are further divided into 21 possible foreground stars (class ``FRG'') and 175 likely YSO members of the region (classes ``NOD'' and ``DSK''). The ``FRG'' class is composed of stars with low median energies and NIR colors (green in Figure \ref{fig_mem} (a) and Figure \ref{fig_ccds_xray} (a)) that are not listed in the optical catalog of young stars from SA05.

The application of the MIR source classification scheme of \citet{Gutermuth09} to the X-ray YSO sample yields zero extragalactic candidates. The scheme can not be applied to the ``EXG'', ``UNC1'', and ``UNC2'' objects, since many of them lack MIR counterparts and all lack [5.8] and [8.0] band magnitudes.

Only the 175 likely YSO members (classes ``NOD'' and ``DSK'') are used in our analyses of the IC 1396A stellar population. In \S \ref{disk_classes_subsection} these sources are further separated into 124 diskless and 51 disky X-ray YSOs. The $J$ versus $J-H$ color-magnitude diagram in Figure \ref{fig_JvsJH_acis} shows that the X-ray catalog is sensitive to diskless and disky stars down to $J \ga 15$~mag and $J \ga 14$~mag, respectively. X-ray YSOs outside the globule have typical source extinctions of $A_V < 2$~mag, while inside the globule stars can be subject to higher extinction of up to $A_V = 7-10$~mag. The optical catalog of SA05 within the ACIS field (green $\circ$) is not sensitive to stars with $J > 14$~mag or highly-absorbed stars. Figure \ref{fig_JvsJH_disky} shows that at $J \ga 14$~mag the X-ray disky sample is well complemented by the MC09 catalog of disky stars near and inside the globule as well as our custom IRAC catalog of disky stars outside the globule (\S \ref{non_chandra_members_subsection}).

In the ACIS field, previous studies had located 76 H$\alpha$ emitters and Li absorbers (SA05, B11) and/or 69 mid-infrared excess stars (MC09) representing Class I/II YSOs. The Chandra source list (Tables \ref{tbl_xray_photometry}-\ref{tbl_derived_props}) includes 61 of these previously known members, and discovers 114 new mostly diskless members\footnote{Notice on Figure \ref{fig_mem} (a) unusually hard and bright emission from one of the reddest Class~I protostars in the cloud, {\it Chandra} source \#32 (MC09 source IC1396A-7; source $\varepsilon$ from \citet{Reach04}). Detailed X-ray spectral and variability analyses from IC1396A-7 are out of the scope of this paper and will be presented elsewhere.}.

\begin{figure*}
\centering
\includegraphics[angle=0.,width=180mm]{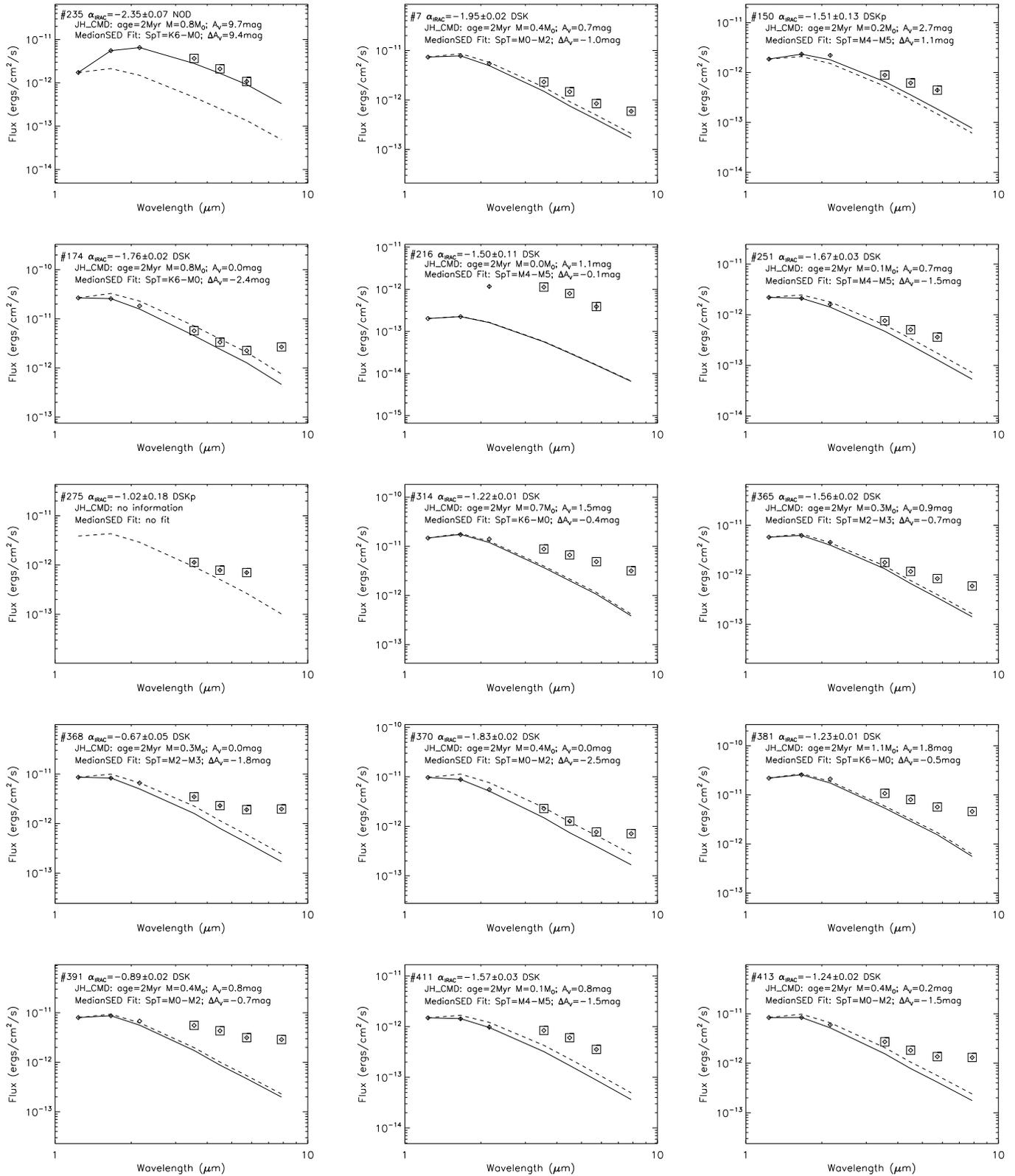}
\caption{IR SEDs for 14 new X-ray disky stars that are not listed in the previous YSO catalogs. Plus, a diskfree SED for the highly absorbed X-ray source \# 235. $JHK_s$ (diamond) and IRAC-band (square) flux points with usually small errors. The dashed and solid lines give the original and (de) reddened IC 348 median SED from \citet{Lada06} fitted to the data. The top two lines of the panel legends give information on stellar mass and source extinction estimates derived from the color-magnitude $J$ versus $J-H$ diagram. The third line gives the spectral class of the corresponding IC 348 median SED template from \citet{Lada06}  and reddening applied to the original template SED to fit the observed IC 1396A source SED. \label{fig_seds_Xray}}
\end{figure*}

\subsection{Pre-Main-Sequence Disk Classification of X-ray stars}\label{disk_classes_subsection}

We base the evolutionary classification of X-ray emitting YSOs in the IC 1396A region on a comparison of their IR spectral energy distributions (SEDs) with the SEDs of YSOs in the well-studied IC~348 cluster in the Perseus molecular cloud \citep{Lada06}. Specifically, we compare the observed IC 1396A SEDs in $2MASS+$IRAC IR bands to the (de)reddened median SED templates of IC~348 PMS stellar photospheres. This is our primary classification method and is described in detail in \citet[][their \S 3]{Getman09}. We further confirm these classifications through calculations of the observed SED slope $\alpha_0$,  and locations in IRAC color-color diagrams. Results of our disk classification are given in Table~\ref{tbl_derived_props}.

In the SED-based classification, rough estimates of individual stellar masses of the IC 1396A X-ray YSOs are obtained from the $J$ versus $J-H$ color-magnitude diagram (Figure \ref{fig_JvsJH_acis}) using a set of trial age values (1, 2, 3, 4, and 10~Myr). For each trial age, an IC~348 template SED at the corresponding spectral sub-class is assigned to each of the X-ray sources. The IC~348 template is (de)reddened and normalized to match both $J$ and $H$-band fluxes of the observed IC 1396A source's SED. The reddening applied to the original IC~348 templates to fit the observed IC~1396A SEDs ($\Delta A_V$ in Figure \ref{fig_seds_Xray}) summed with the median source extinction for the diskless YSO sample in IC~348 ($A_V = 1.6$~mag from Table 2 in \citet{Lada06}) gives an estimate of the source extinction from the SED fitting, $A_{V,SED}$. We find that $A_{V,SED}$ is in a reasonable agreement with the source extinction derived from the $J$ versus $J-H$ color-magnitude diagram, $A_{V,CMD}$. For example, for the trial age of $2$~Myr the median, mean, and standard deviation of the $A_{V,SED} - A_{V,CMD}$ difference are $-0.1$, $-0.1$, and $0.6$~mag, respectively.

Visual examination of deviation of the observed IC 1396A SED from the corresponding matched IC~348 photospheric template allows disk classification of IC 1396A X-ray YSOs. While inferred values for individual stellar masses strongly depend on the choice of age, the final classification discriminating diskless and disky stars is not age sensitive. With regard to the 28 X-ray YSOs with MC09 counterparts, our disk classification agrees with that of MC09 for all but one ($Chandra$ source \#235 = MC09 source IC1396A-46) sources.  Disky SEDs for 14 X-ray YSOs that do not have counterparts in any previous stellar catalog and a diskfree SED for the highly absorbed X-ray YSO \#235 are shown in Figure \ref{fig_seds_Xray}.

\begin{figure}
\centering
\includegraphics[angle=0.,width=78mm]{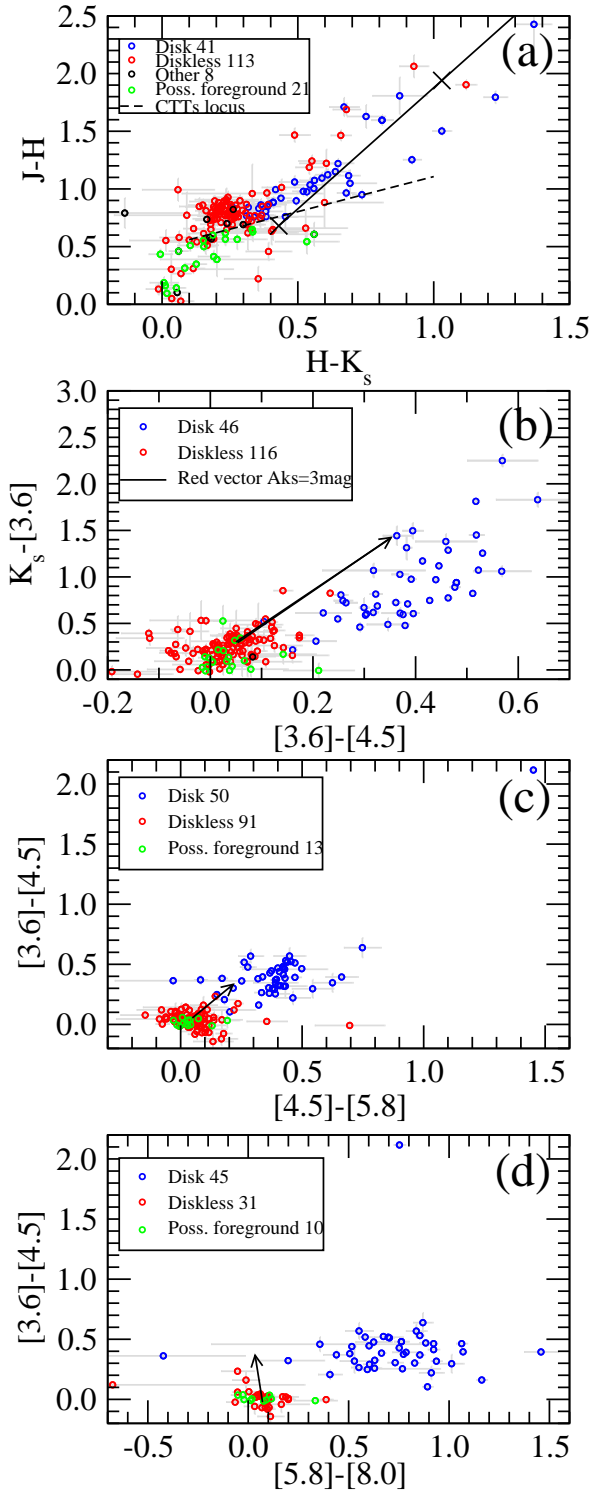}
\caption{IR color-color diagrams of X-ray sources. For all panels, YSO diskfree (Class~III) stars with available IR photometry information are in red, YSO disky (Class~I and II) stars are in blue, possible foreground stars unrelated to the region are in green. On panels (b)-(d) the reddening vectors show $A_K \sim 2$~mag using the extinction law from \citet{Flaherty07}.  (a) $J-H$ versus $H-K_s$. The dashed line indicates the classical T Tauri star locus \citep{Meyer97}. The solid line is a reddening vector marked at intervals of $A_V = 10$~mag.  (b) $K_s - [3.6]$ versus $[3.6] - [4.5]$. (c) $[3.6]-[4.5]$ versus $[4.5] - [5.8]$. (d) $[3.6] - [4.5]$ versus $[5.8] - [8.0]$. \label{fig_ccds_xray}}
\end{figure}

\begin{figure}
\centering
\includegraphics[angle=0.,width=85mm]{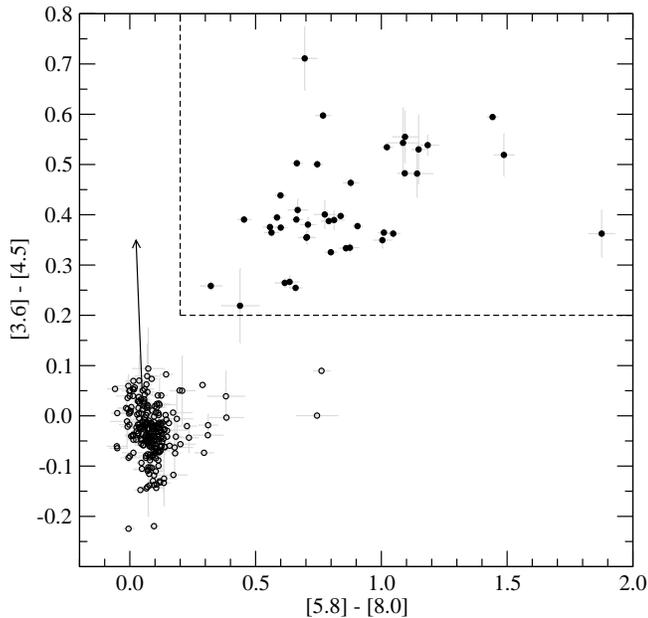}
\caption{IR color-color diagram for non-{\it Chandra} sources with reliable photometry identified through the analysis of {\it Spitzer}-IRAC data in the current work (\S \ref{non_chandra_members_subsection}). Classification criteria for selecting disky YSOs, $[3.6] - [4.5] \ga 0.2$~mag and $[5.8] - [8.0] \ga 0.2$~mag, are shown by the dashed lines. Forty two non-{\it Chandra} YSOs additional to the disky stars from MC09 are identified (filled circles). The reddening vector shows $A_K \sim 2$~mag using the extinction law from \citet{Flaherty07}. \label{fig_cmd_nonXray_disky}}
\end{figure}

We also performed least-squares linear fits to the $\log(\lambda F_{\lambda})$ values in the four IRAC wavelength band to obtain the observed (not de-reddened) SED spectral index, $\alpha_{0}$.  These are tabulated in column~8 of Table~\ref{tbl_derived_props}. As in the case of the X-ray selected YSOs in the Cepheus star forming region \citep[][their Figure 6]{Getman09}, the distribution of the spectral index for IC 1396A X-ray YSOs has a bimodal shape with peaks at $\alpha_{0} \sim -1.2$ and $-2.7$ corresponding to diskbearing and diskless star samples, respectively (graph is not shown). In agreement with the SED-based classification, all but one diskless X-ray stars have SED spectral indices of $\alpha_{0} < -2.1$, and all disky X-ray stars have indices of $\alpha_{0} > -2.1$.

In Figure \ref{fig_ccds_xray} we compare the SED-based classifications of X-ray stars with the expected loci of YSO stars in NIR and MIR color-color diagrams \citep[e.g.][]{Allen04, Megeath04, Hartmann05}. The positions of stars in the color-color diagrams are in good agreement with our previous classification of stars --- simple color criteria that disky stars have $[3.6] - [4.5] \ga 0.2$, $[4.5] - [5.8] \ga 0.2$, or $[5.8] - [8.0] \ga 0.2$, select nearly all of the SED-based classified disky stars. A couple of stars with the $3.6-4.5$ color near zero and with an excess at longer wavelengths could be transition disk objects, candidates to have inner holes in their disks rather than being diskless.

\section{NON-{\it CHANDRA} INFRARED-EXCESS MEMBERS}\label{non_chandra_members_subsection}

For a given mass, the X-ray detection efficiency of Class~II stars is somewhat lower than that of Class~III stars \citep[][and Figure \ref{fig_JvsJH_acis} here]{Getman09,Stelzer11}. To compensate for this effect, this section presents additional non-{\it Chandra} IR-excess selected disky stars. The MIR catalog from MC09 provides an excellent (\S \ref{imf_analysis_globule_subsection}) selection of disky members near/inside the globule. The catalog also includes all disky stars near/inside the globule reported by \citet{Reach04} and \citet{Sicilia-Aguilar06}. However, MC09 covers only one fifth of the ACIS field. Thus, we use the results from our {\it Spitzer}-IRAC data analysis (\S \ref{irac_data_reduction_subsection}) to produce an additional sample of non-{\it Chandra} disky stars for the remaining ACIS field and beyond.

\begin{table*}\tiny
\centering
 \begin{minipage}{180mm}
 \caption{Non-{\it Chandra} disky stars additional to MC09 stars. Column 1: Source number for non-{\it Chandra} IR-excess sources found in our IRAC data analysis. These sources are located around the globule and are complementary to the IR-excess sources in/near the globule cataloged by MC09. Columns 2 and 3: IRAC right ascension and declination for epoch J2000.0 in degrees. Columns 4-7: $2MASS$ $JHK_s$ magnitudes, and $2MASS$ photometry quality and confusion-contamination flag. Columns 8-12: IRAC magnitudes and the aperture flag derived in this work. The flag gives photometric apertures and level of source contamination from nearby sources and nebular IR emission: 2-pixel (1.73$\arcsec$), 3-pixel (2.6$\arcsec$), and 4-pixel (3.46$\arcsec$) apertures with contaminating flux from a neighboring source of no more than $10$\%, $5$\%, and $5$\% in [3.6] band, respectively. Column 13: Apparent SED slope from IRAC photometry with $1\sigma$ error. Column 14: A positional flag indicating if a source lies within the ACIS field of view (``1'') or not (``0''). Columns 15 and 16: Stellar counterparts from B11 and SA05.}
 \label{tbl_nonXray_disky}
 \begin{tabular}{@{\vline}|c@{\vline}|c@{\vline}c@{\vline}c@{\vline}c@{\vline}c@{\vline}c@{\vline}c@{\vline}c@{\vline}c@{\vline}c@{\vline}c@{\vline}c@{\vline}c@{\vline}c@{\vline}c@{\vline}}
\cline{1-16}
&&&&&&&&&&&&&&&\\

No. & R.A. & Decl. &
$J$ & $H$ & $K_s$ &
F1 & [3.6] &
[4.5] & [5.8] &
[8.0] & F2 &
$\alpha_0$ & F3 & B11 & SA05\\
&(deg)&(deg)&(mag)&(mag)&(mag)&&(mag)&(mag)&(mag)&(mag)&&&&&\\
(1)&(2)&(3)&(4)&(5)&(6)&(7)&(8)&(9)&(10)&(11)&(12)&(13)&(14)&(15)&(16)\\
&&&&&&&&&&&&&&&\\
\cline{1-16}
&&&&&&&&&&&&&&&\\

1 & 323.880500 &  57.524333 &  $  14.53 \pm ... $  &  $  13.63 \pm   0.05 $  &  $  13.22 \pm   0.04 $  & UAA0c0 &  $  12.41 \pm   0.01 $  &  $  12.15 \pm   0.01 $  &  $  12.06 \pm   0.02 $  &  $  11.73 \pm   0.04 $  & 4 &  $ -2.04 \pm  0.02 $  & 0 &  19 & ...\\
 2 & 323.887000 &  57.533194 &  $  14.64 \pm   0.04 $  &  $  13.81 \pm   0.04 $  &  $  13.51 \pm   0.04 $  & AAA000 &  $  13.00 \pm   0.01 $  &  $  12.64 \pm   0.00 $  &  $  12.16 \pm   0.01 $  &  $  11.11 \pm   0.02 $  & 4 &  $ -1.09 \pm  0.02 $  & 1 &  21 & ...\\
 3 & 324.243333 &  57.650250 &  $  16.48 \pm   0.15 $  &  $  15.40 \pm   0.11 $  &  $  14.62 \pm   0.09 $  & BBAccc &  $  13.79 \pm   0.03 $  &  $  13.23 \pm   0.04 $  &  $  12.66 \pm   0.04 $  &  $  11.57 \pm   0.03 $  & 2 &  $ -0.27 \pm  0.05 $  & 1 & ... & ...\\
 4 & 324.244708 &  57.646667 &  $  14.14 \pm   0.03 $  &  $  13.34 \pm   0.04 $  &  $  12.97 \pm   0.03 $  & AAA000 &  $  12.46 \pm   0.00 $  &  $  12.21 \pm   0.00 $  &  $  11.90 \pm   0.01 $  &  $  11.24 \pm   0.02 $  & 4 &  $ -1.71 \pm  0.01 $  & 1 & ... & ...\\
 5 & 324.246292 &  57.651556 &  $  13.57 \pm   0.02 $  &  $  12.65 \pm   0.03 $  &  $  12.26 \pm   0.03 $  & AAA000 &  $  11.60 \pm   0.00 $  &  $  11.12 \pm   0.00 $  &  $  10.57 \pm   0.01 $  &  $   9.48 \pm   0.03 $  & 4 &  $ -0.84 \pm  0.01 $  & 1 &  34 & ...\\
 6 & 324.283917 &  57.604472 &  $  14.58 \pm   0.04 $  &  $  13.65 \pm   0.04 $  &  $  13.27 \pm   0.04 $  & AAA000 &  $  12.60 \pm   0.00 $  &  $  12.21 \pm   0.00 $  &  $  11.84 \pm   0.01 $  &  $  11.05 \pm   0.03 $  & 4 &  $ -1.30 \pm  0.02 $  & 1 & ... & ...\\
 7 & 324.300708 &  57.457306 &  $  13.39 \pm   0.03 $  &  $  12.21 \pm   0.03 $  &  $  11.52 \pm   0.03 $  & AAA000 &  $  10.64 \pm   0.00 $  &  $  10.25 \pm   0.00 $  &  $   9.96 \pm   0.01 $  &  $   9.37 \pm   0.01 $  & 4 &  $ -1.42 \pm  0.01 $  & 1 & ... & 11-2131\\
 8 & 324.307917 &  57.457528 &  $  14.27 \pm   0.04 $  &  $  13.15 \pm   0.04 $  &  $  12.40 \pm   0.03 $  & AAAc00 &  $  11.42 \pm   0.01 $  &  $  10.92 \pm   0.00 $  &  $  10.52 \pm   0.01 $  &  $   9.78 \pm   0.01 $  & 4 &  $ -1.01 \pm  0.02 $  & 1 &  40 & ...\\
 9 & 324.309125 &  57.604889 &  $  14.58 \pm   0.03 $  &  $  13.59 \pm   0.04 $  &  $  13.01 \pm   0.03 $  & AAA000 &  $  11.91 \pm   0.00 $  &  $  11.38 \pm   0.00 $  &  $  10.95 \pm   0.01 $  &  $   9.93 \pm   0.02 $  & 4 &  $ -0.79 \pm  0.01 $  & 1 & ... & ...\\
10 & 324.314375 &  57.454722 &  $  14.46 \pm   0.04 $  &  $  13.48 \pm   0.04 $  &  $  12.95 \pm   0.04 $  & AAA000 &  $  12.14 \pm   0.01 $  &  $  11.78 \pm   0.00 $  &  $  11.42 \pm   0.01 $  &  $  10.86 \pm   0.03 $  & 4 &  $ -1.44 \pm  0.02 $  & 1 & ... & ...\\
11 & 324.321458 &  57.479806 &  $  15.02 \pm   0.05 $  &  $  14.20 \pm   0.05 $  &  $  13.93 \pm   0.07 $  & AAA000 &  $  13.31 \pm   0.01 $  &  $  12.96 \pm   0.01 $  &  $  12.55 \pm   0.03 $  &  $  11.55 \pm   0.02 $  & 3 &  $ -0.93 \pm  0.03 $  & 1 &  42 & ...\\
12 & 324.333417 &  57.492861 &  $  14.88 \pm   0.04 $  &  $  14.11 \pm   0.06 $  &  $  13.79 \pm   0.06 $  & AAA000 &  $  13.14 \pm   0.01 $  &  $  12.88 \pm   0.01 $  &  $  12.58 \pm   0.02 $  &  $  11.94 \pm   0.03 $  & 4 &  $ -1.66 \pm  0.03 $  & 1 & ... & ...\\
13 & 324.339625 &  57.436778 &  $  14.23 \pm   0.07 $  &  $  13.36 \pm   0.07 $  &  $  12.89 \pm   0.04 $  & AAAcc0 &  $  12.40 \pm   0.04 $  &  $  11.91 \pm   0.03 $  &  $  11.57 \pm   0.06 $  &  $  10.43 \pm   0.02 $  & 2 &  $ -0.51 \pm  0.04 $  & 1 & ... & ...\\
14 & 324.350333 &  57.403167 &  $  14.20 \pm   0.07 $  &  $  13.29 \pm   0.08 $  &  $  12.82 \pm   0.05 $  & AAAccc &  $  11.94 \pm   0.04 $  &  $  11.41 \pm   0.05 $  &  $  10.84 \pm   0.02 $  &  $   9.69 \pm   0.02 $  & 2 &  $ -0.12 \pm  0.05 $  & 1 &  44 & ...\\
15 & 324.351917 &  57.526639 &  $  14.54 \pm   0.03 $  &  $  13.79 \pm   0.05 $  &  $  13.28 \pm   0.04 $  & AAA000 &  $  12.59 \pm   0.00 $  &  $  12.22 \pm   0.00 $  &  $  11.88 \pm   0.01 $  &  $  10.87 \pm   0.02 $  & 4 &  $ -1.22 \pm  0.02 $  & 1 & ... & 21372447+5731359\\
16 & 324.364167 &  57.523222 &  $  14.38 \pm   0.04 $  &  $  13.34 \pm   0.04 $  &  $  12.98 \pm   0.03 $  & AAAc00 &  $  12.27 \pm   0.05 $  &  $  12.05 \pm   0.06 $  &  $  11.58 \pm   0.04 $  &  $  11.15 \pm   0.06 $  & 2 &  $ -1.52 \pm  0.08 $  & 1 & ... & ...\\
17 & 324.370583 &  57.601222 &  $  13.99 \pm   0.03 $  &  $  13.03 \pm   0.03 $  &  $  12.63 \pm   0.03 $  & AAA000 &  $  11.86 \pm   0.00 $  &  $  11.47 \pm   0.00 $  &  $  11.09 \pm   0.01 $  &  $  10.43 \pm   0.02 $  & 4 &  $ -1.33 \pm  0.01 $  & 1 & ... & 14-1017\\
18 & 324.392083 &  57.575333 &  $  15.00 \pm   0.04 $  &  $  13.98 \pm   0.04 $  &  $  13.49 \pm   0.04 $  & AAA000 &  $  12.64 \pm   0.00 $  &  $  12.24 \pm   0.00 $  &  $  11.91 \pm   0.01 $  &  $  11.45 \pm   0.02 $  & 4 &  $ -1.44 \pm  0.02 $  & 1 & ... & ...\\
19 & 324.398792 &  57.549556 &  $  14.86 \pm   0.04 $  &  $  14.02 \pm   0.05 $  &  $  13.62 \pm   0.06 $  & AAAc00 &  $  13.21 \pm   0.01 $  &  $  12.81 \pm   0.02 $  &  $  12.45 \pm   0.04 $  &  $  11.68 \pm   0.02 $  & 3 &  $ -1.11 \pm  0.03 $  & 1 &  48 & ...\\
20 & 324.418667 &  57.575889 &  $  14.99 \pm   0.05 $  &  $  14.09 \pm   0.06 $  &  $  13.77 \pm   0.07 $  & AAA0c0 &  $  13.13 \pm   0.03 $  &  $  12.61 \pm   0.03 $  &  $  11.95 \pm   0.03 $  &  $  10.47 \pm   0.02 $  & 2 &  $  0.34 \pm  0.04 $  & 1 & ... & ...\\
21 & 324.428667 &  57.579500 &  $  14.96 \pm   0.04 $  &  $  14.18 \pm   0.05 $  &  $  13.76 \pm   0.05 $  & AAAc00 &  $  13.15 \pm   0.01 $  &  $  12.79 \pm   0.01 $  &  $  12.32 \pm   0.02 $  &  $  11.62 \pm   0.03 $  & 4 &  $ -1.22 \pm  0.03 $  & 1 &  51 & ...\\
22 & 324.428792 &  57.608750 &  $  10.57 \pm   0.02 $  &  $  10.10 \pm   0.03 $  &  $   9.84 \pm   0.02 $  & AAA000 &  $   8.95 \pm   0.00 $  &  $   8.35 \pm   0.00 $  &  $   7.59 \pm   0.01 $  &  $   6.15 \pm   0.02 $  & 4 &  $ -0.27 \pm  0.01 $  & 1 & ... & ...\\
23 & 324.442167 &  57.574444 &  $  14.92 \pm   0.05 $  &  $  14.03 \pm   0.05 $  &  $  13.64 \pm   0.06 $  & AAA000 &  $  12.92 \pm   0.00 $  &  $  12.56 \pm   0.00 $  &  $  12.20 \pm   0.02 $  &  $  11.50 \pm   0.01 $  & 4 &  $ -1.30 \pm  0.02 $  & 1 &  53 & ...\\
24 & 324.445292 &  57.532278 &  $  14.85 \pm   0.05 $  &  $  13.95 \pm   0.04 $  &  $  13.63 \pm   0.05 $  & AAA000 &  $  12.88 \pm   0.00 $  &  $  12.55 \pm   0.00 $  &  $  12.21 \pm   0.02 $  &  $  11.35 \pm   0.02 $  & 4 &  $ -1.36 \pm  0.02 $  & 1 & ... & ...\\
25 & 324.449833 &  57.545083 &  $  14.15 \pm   0.04 $  &  $  13.14 \pm   0.03 $  &  $  12.77 \pm   0.04 $  & AAA000 &  $  12.03 \pm   0.01 $  &  $  11.69 \pm   0.01 $  &  $  11.36 \pm   0.01 $  &  $  10.49 \pm   0.03 $  & 4 &  $ -1.47 \pm  0.02 $  & 1 &  54 & ...\\
26 & 324.452208 &  57.499667 &  $  15.04 \pm   0.06 $  &  $  14.31 \pm   0.06 $  &  $  14.05 \pm   0.07 $  & AAA000 &  $  13.19 \pm   0.01 $  &  $  12.78 \pm   0.02 $  &  $  12.37 \pm   0.03 $  &  $  11.71 \pm   0.02 $  & 3 &  $ -1.16 \pm  0.03 $  & 1 & ... & ...\\
27 & 324.452292 &  57.405056 &  ...  &  ...  &  ...  & ... &  $  13.68 \pm   0.00 $  &  $  13.08 \pm   0.00 $  &  $  12.59 \pm   0.02 $  &  $  11.82 \pm   0.02 $  & 4 &  $ -0.67 \pm  0.02 $  & 0 & ... & ...\\
28 & 324.453917 &  57.389111 &  $  14.66 \pm   0.06 $  &  $  13.68 \pm   0.06 $  &  $  13.25 \pm   0.05 $  & AAAc00 &  $  12.37 \pm   0.01 $  &  $  11.98 \pm   0.01 $  &  $  11.64 \pm   0.02 $  &  $  10.93 \pm   0.02 $  & 3 &  $ -1.28 \pm  0.02 $  & 0 &  55 & ...\\
29 & 324.454958 &  57.464278 &  $  16.04 \pm   0.12 $  &  $  15.23 \pm   0.12 $  &  $  14.30 \pm ... $  & BBUc00 &  $  14.26 \pm   0.03 $  &  $  13.90 \pm   0.03 $  &  $  13.33 \pm   0.05 $  &  $  11.45 \pm   0.02 $  & 3 &  $  0.68 \pm  0.04 $  & 1 & ... & ...\\
30 & 324.455958 &  57.564333 &  $  16.39 \pm   0.14 $  &  $  15.04 \pm   0.10 $  &  $  14.28 \pm   0.09 $  & BAA000 &  $  13.79 \pm   0.03 $  &  $  13.24 \pm   0.06 $  &  $  12.57 \pm   0.06 $  &  $  11.48 \pm   0.03 $  & 2 &  $ -0.17 \pm  0.05 $  & 1 & ... & ...\\
31 & 324.463292 &  57.410000 &  $  15.48 \pm   0.07 $  &  $  14.68 \pm   0.10 $  &  $  14.24 \pm   0.07 $  & AAA000 &  $  13.34 \pm   0.01 $  &  $  12.87 \pm   0.01 $  &  $  12.37 \pm   0.01 $  &  $  11.49 \pm   0.03 $  & 4 &  $ -0.90 \pm  0.02 $  & 0 & ... & ...\\
32 & 324.468958 &  57.432250 &  $  15.21 \pm   0.06 $  &  $  14.42 \pm   0.07 $  &  $  13.97 \pm   0.06 $  & AAA000 &  $  13.72 \pm   0.01 $  &  $  13.18 \pm   0.01 $  &  $  12.56 \pm   0.03 $  &  $  11.38 \pm   0.03 $  & 3 &  $ -0.23 \pm  0.03 $  & 0 & ... & ...\\
33 & 324.496833 &  57.604472 &  $  13.32 \pm   0.03 $  &  $  12.40 \pm   0.03 $  &  $  11.89 \pm   0.02 $  & AAA000 &  $  11.13 \pm   0.00 $  &  $  10.73 \pm   0.00 $  &  $  10.36 \pm   0.01 $  &  $   9.52 \pm   0.01 $  & 4 &  $ -1.21 \pm  0.01 $  & 0 &  57 & 13-1238\\
34 & 324.514542 &  57.693028 &  $  12.50 \pm   0.03 $  &  $  11.48 \pm   0.03 $  &  $  10.85 \pm   0.02 $  & AAA000 &  $   9.83 \pm   0.00 $  &  $   9.51 \pm   0.00 $  &  $   9.24 \pm   0.01 $  &  $   8.44 \pm   0.02 $  & 4 &  $ -1.59 \pm  0.01 $  & 0 & ... & 82-272\\
35 & 324.535125 &  57.446583 &  $  10.51 \pm   0.03 $  &  $   9.72 \pm   0.03 $  &  $   8.77 \pm   0.02 $  & AAA000 &  $   7.30 \pm   0.00 $  &  $   6.80 \pm   0.00 $  &  $   6.41 \pm   0.01 $  &  $   5.75 \pm   0.02 $  & 4 &  $ -1.03 \pm  0.01 $  & 0 & ... & MVA-426\\
36 & 324.535625 &  57.618778 &  $  14.31 \pm   0.03 $  &  $  13.43 \pm   0.03 $  &  $  12.90 \pm   0.03 $  & AAA000 &  $  11.45 \pm   0.00 $  &  $  11.01 \pm   0.00 $  &  $  10.78 \pm   0.01 $  &  $  10.18 \pm   0.02 $  & 4 &  $ -1.37 \pm  0.01 $  & 0 &  58 & 13-1426\\
37 & 324.538667 &  57.557250 &  $  12.39 \pm   0.03 $  &  $  11.57 \pm   0.03 $  &  $  11.20 \pm   0.02 $  & AAA000 &  $  10.60 \pm   0.00 $  &  $  10.22 \pm   0.00 $  &  $   9.85 \pm   0.01 $  &  $   8.95 \pm   0.01 $  & 4 &  $ -1.15 \pm  0.01 $  & 1 & ... & 13-669\\
38 & 324.542417 &  57.452333 &  $  15.05 \pm   0.08 $  &  $  13.83 \pm ... $  &  $  13.63 \pm ... $  & AUUc00 &  $  13.10 \pm   0.06 $  &  $  12.39 \pm   0.01 $  &  $  12.30 \pm   0.04 $  &  $  11.61 \pm   0.02 $  & 2 &  $ -1.49 \pm  0.04 $  & 0 & ... & ...\\
39 & 324.550000 &  57.416861 &  $  14.53 \pm   0.05 $  &  $  13.58 \pm   0.04 $  &  $  13.32 \pm   0.05 $  & AAA000 &  $  12.66 \pm   0.01 $  &  $  12.40 \pm   0.01 $  &  $  12.10 \pm   0.02 $  &  $  11.48 \pm   0.03 $  & 4 &  $ -1.71 \pm  0.03 $  & 0 & ... & ...\\
40 & 324.594958 &  57.671500 &  $  14.83 \pm   0.04 $  &  $  13.78 \pm   0.04 $  &  $  13.35 \pm   0.03 $  & AAA000 &  $  12.36 \pm   0.00 $  &  $  11.98 \pm   0.00 $  &  $  11.74 \pm   0.01 $  &  $  11.19 \pm   0.02 $  & 4 &  $ -1.52 \pm  0.02 $  & 0 &  64 & ...\\
41 & 324.614250 &  57.518917 &  $  12.36 \pm   0.04 $  &  $  11.36 \pm   0.04 $  &  $  10.77 \pm   0.02 $  & AAA000 &  $  10.02 \pm   0.00 $  &  $   9.65 \pm   0.00 $  &  $   9.33 \pm   0.01 $  &  $   8.73 \pm   0.02 $  & 4 &  $ -1.44 \pm  0.01 $  & 0 &  70 & 13-236\\
42 & 324.616833 &  57.512889 &  $  12.91 \pm   0.03 $  &  $  11.98 \pm   0.04 $  &  $  11.25 \pm   0.05 $  & AAA000 &  $  10.19 \pm   0.01 $  &  $   9.80 \pm   0.02 $  &  $   9.62 \pm   0.03 $  &  $   8.81 \pm   0.02 $  & 3 &  $ -1.32 \pm  0.03 $  & 0 & ... & 13-157\\
&&&&&&&&&&&&&&&\\
\cline{1-16}
\end{tabular}
\end{minipage}
\end{table*}

Our IRAC data analysis finds 292 sources not detected by {\it Chandra} and not published by MC09, with photometric errors $<0.1$~mag in all four IRAC bands. Figure \ref{fig_cmd_nonXray_disky} shows the $[3.6] - [4.5]$ versus $[5.8] - [8.0]$ diagram for these sources. Using the IR color-color classification criteria presented in \S \ref{disk_classes_subsection}, we identify 42 (out of 292) IR-excess stellar candidates. According to the classification scheme of \citet{Gutermuth09}, only one of these sources, source \# 29, could be an extragalactic contaminant. Since the IC~1396 region is in the outer Galaxy, we do not expect the sample to be contaminated by asymptotic giant branch or classical Be stars. 

\begin{figure*}
\centering
\includegraphics[angle=0.,width=160mm]{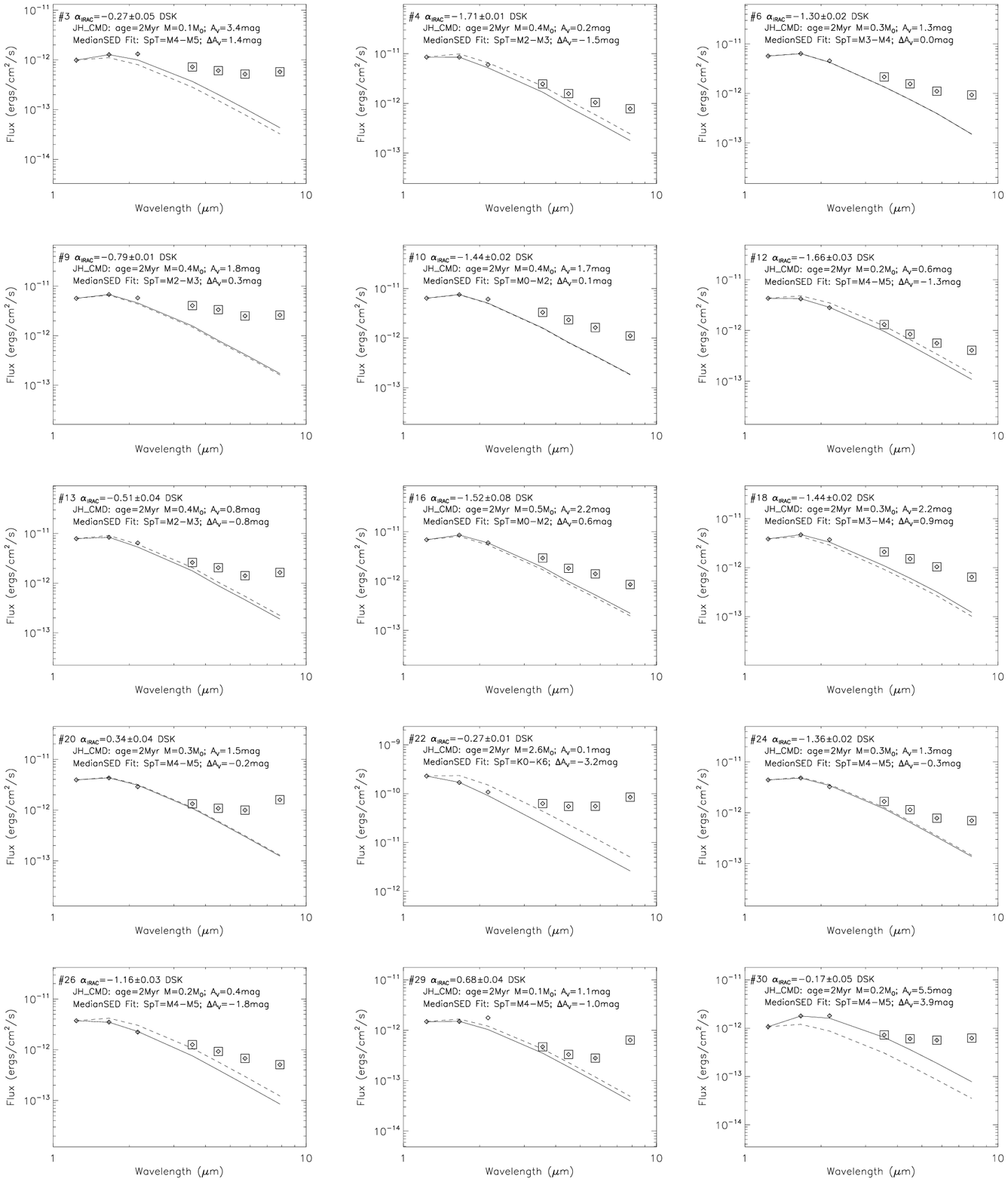}
\caption{IR SEDs for 15 new non-{\it Chandra} disky YSOs within the ACIS field that are not in the source catalogs of SA05, MC09, or B11. See Figure \ref{fig_seds_Xray} for details. \label{fig_seds_nonXray_disky}}
\end{figure*}

Table \ref{tbl_nonXray_disky} reports their NIR and MIR magnitudes, apparent SED slope from IRAC photometry, a flag indicating if a source lies within the ACIS field, and stellar counterparts from SA05 and B11. Out of 42 stars, 20 (without SA05 and B11 counterparts) are newly discovered members with 15 of them found within the ACIS field. IR SEDs for these 15 sources are given in Figure \ref{fig_seds_nonXray_disky}. They deviate from photospheric SEDs at wavelengths longer than at least 3.6 $\mu$m. As expected, the majority of these new non-{\it Chandra} disky YSOs are fainter than the X-ray selected disky sample (compare cyan $\sq$s and blue $\times$s in Figure \ref{fig_JvsJH_disky}).

\begin{figure*}
\centering
\includegraphics[angle=0.,width=160mm]{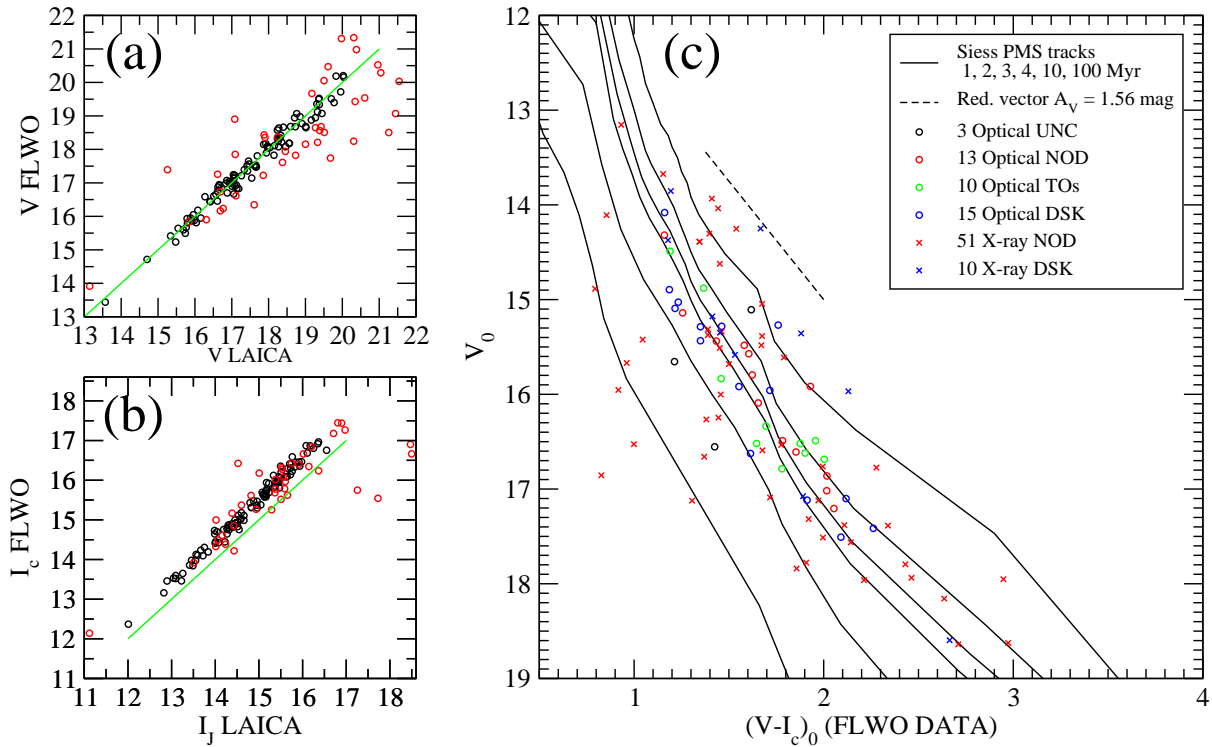}
\caption{Derivation of ages. Comparison of $V$-band (a) and $I$-band (b) magnitudes for young SA05/SA10 and X-ray stars that have both FLWO and LAICA photometry. Black (red) circles mark 102 (40) stars that have $V$, $I$-band magnitudes consistent (inconsistent) between the FLWO and LAICA observations. Unity lines are shown in green. (c) $VI_c$ color-magnitude diagram for the 102 stars with consistent photometry: 61 X-ray selected stars ($\times$) and 41 SA05/SA10 stars without X-ray counterparts ($\circ$). Colors indicate different disk classes: diskfree (red), diskbearing (blue), transition disk objects (green), and uncertain class (black). Magnitudes and colors have been corrected for the average extinction of $<A_V> \sim 1.56$~mag (SA05). The black solid lines show PMS isochrones for ages 1, 2, 3, 4, 10, 100~Myr from evolutionary tracks of \citet{Siess00} assuming distance of 870~pc \citep{Contreras02}. The black dashed line shows a reddening vector of $A_V = 1.56$~mag. \label{fig_age_derivation}}
\end{figure*}

\section{STELLAR AGE ESTIMATES}\label{age_analysis_section}

Indications of an age gradient among PMS stars, decreasing from the central Tr 37 cluster towards the IC 1396A globule, have been reported by SA05 (see their Figure 11) and B11 (their Figure 17). We seek here to confirm this trend and find out if it is statistically significant by taking advantage of the increased stellar sample size near/around the globule and the availability of multi-epoch optical data for many sample members. This requires deriving and comparing in a homogeneous way stellar ages of the combined optical SA05/SA10 and X-ray selected stellar sample.

It is well known that estimates of individual stellar ages can be very uncertain due to several possible causes such as measurement uncertainty, stellar variability, binarity, different accretional histories, extinction uncertainty, model uncertainty, and distance uncertainty \citep[e.g.][]{Baraffe09, Soderblom10, Preibisch12}. However, here we are not so much interested in precise ages for individual objects as in the general trends of age distributions and consistent treatment for comparison between different stellar populations. In spite of all the uncertainties, the methods of isochrones and color-magnitude diagrams often allow discrimination among YSO regions of different characteristic ages \citep[e.g.][]{Burningham05, Mayne07}. The isochronal ages were shown to be in agreement with gravity related indices \citep{Lawson09}. Similar isochronal ages found for binary components in the Taurus-Auriga region are in agreement with the commonly assumed coevality of binary systems \citep{Kraus09}.

First, we clean our source sample to reduce some of the effects that cause age uncertainty. Our initial sample is limited to the stars that have both FLWO and LAICA photometric measurements. These are 89 X-ray stars and 53 SA05-SA10 optical stars without X-ray counterparts. Figure \ref{fig_age_derivation}~(a),(b) compares their $VI$-band magnitudes. The systematic difference between the $I_C$ and $I_J$-band magnitudes is due to the different photometric systems, FLWO-Cousins versus LAICA-Johnson. The linear transformation $I_C = 0.2954 + 1.0138 \times I_J$ minimizes the differences for all but a few sources with $I_J>17$~mag. After this $I$-band correction, large spreads in $\Delta V$ and $\Delta I$ measurements are seen. The spreads can be mainly attributed to observational uncertainty and stellar variability. To suppress these effects, the sample is truncated to remove sources with $\vert \Delta V \vert > 0.4$~mag and $\vert \Delta I \vert > 0.4$~mag. The sample is further reduced by removing several sources with possibly high extinction ($J-H>1.2$~mag corresponding to $A_V>3$~mag) and/or with known spectral types G-K1. For the latter, their measured ages could be overestimated by incorrect consideration of birth line effects \citep{Hartmann03}. Our final source sample comprises 102 stars (black points in Figure \ref{fig_age_derivation}~(a),(b)). 

The apparent $V$-band magnitudes and $V-I_C$ colors (and transformed $V-I_J$ colors) are then de-reddened by an average extinction of $<A_V> = 1.56$~mag (SA05,B11) using the reddening law of $A_I/A_V = 0.6$ \citep{BessellBrett88}. Comparison with the PMS evolutionary tracks of \citet{Siess00} yields the individual age estimates ($t_{FLWO}$, $t_{LAICA}$) shown in Table~\ref{tbl_derived_props}.

\begin{figure*}
\centering
\includegraphics[angle=0.,width=160mm]{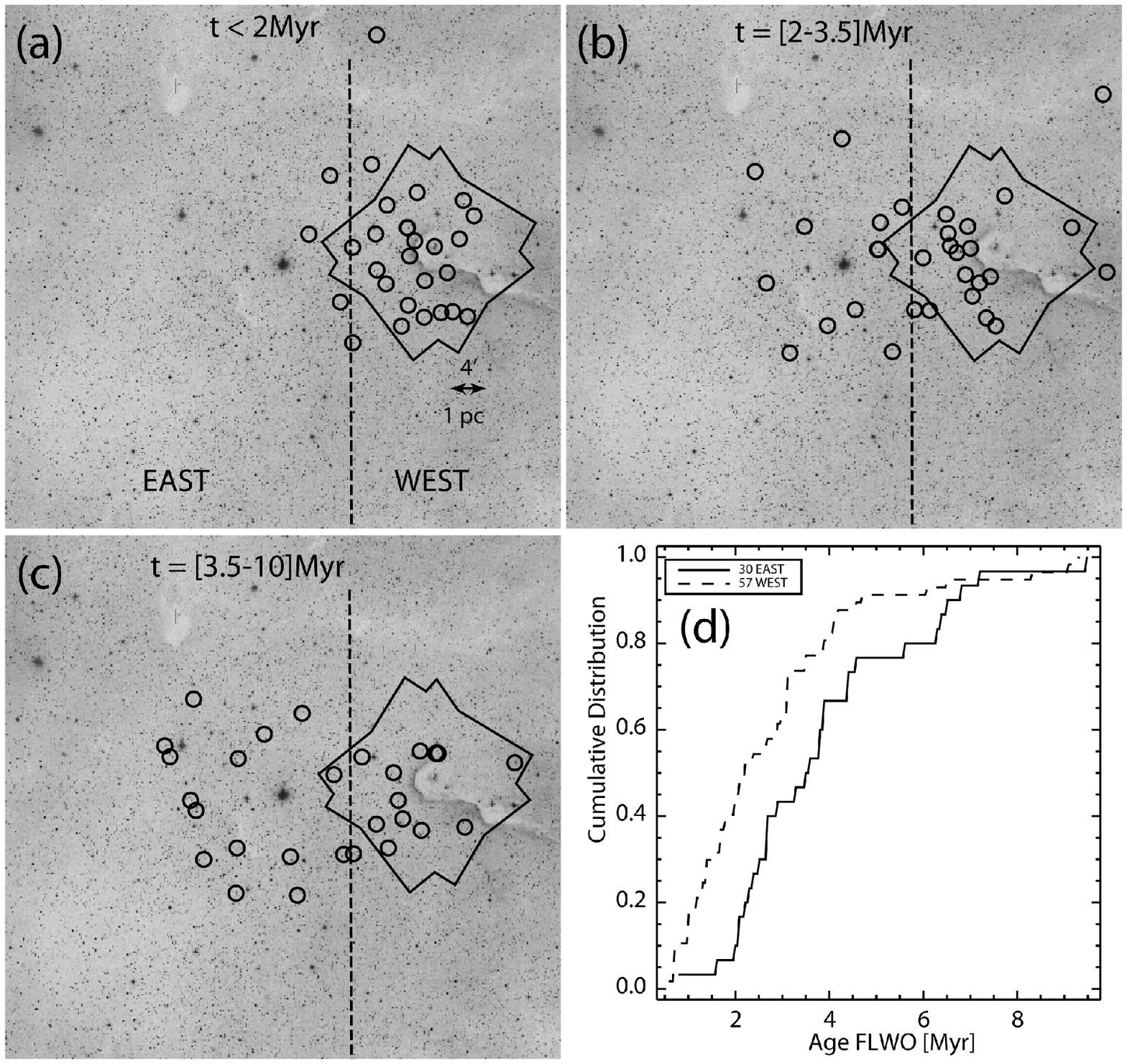}
\caption{Age analysis for the combined optical SA05 and X-ray selected young stellar sample. Age-stratified stellar sub-samples are shown on different panels: $t<2$~Myr (a), $2<t<3.5$~Myr (b), and $3.5<t<10$~Myr (c). The polygon indicates the ACIS field of view. The dashed line at roughly half projected distance between the O6 star HD~206267 and the IC 1396A globule is a demarcation line that divides the region into two sub-regions, East and West. (d) Cumulative distributions of age are compared between the 30 East and 57 West stars with available ages. \label{fig_age_strata}}
\end{figure*}

For the final source sample, statistical photometric uncertainties for both $V$ and $I$-band magnitudes are $<0.04$~mag. However, the systematic uncertainty still prevails --- only 9\% (22\%) of the sources have $\Delta V < 0.04$~mag ($\Delta I < 0.04$~mag), but 68\% (90\%) have $\Delta V < 0.2$~mag ($\Delta I < 0.2$~mag). The FLWO$-$LAICA age differences quantify the effect of variability on the inferred age estimates: $|t_{FLWO}-t_{LAICA}|/t_{FLWO}$ is better than 10\%, 20\%, 30\%, 40\%, 50\%, and 60\% for 27\%, 49\%, 60\%, 70\%, 75\%, and 80\% of the sources, respectively. 

Figure \ref{fig_age_derivation}~(c) shows an optical color-magnitude diagram for the FLWO data (the LAICA data look similar). There are a dozen diskless stars with inferred ages $>10$~Myr. Some of them could be members of a possible old distributed population. Half have inferred ages exceeding 100~Myr. This value seems to be unrealistically high considering that the average age for young stars in the region is around $4$~Myr (SA05). This may point to variability or to misclassified foreground and background stars. For the analysis below, we limit the sample to stars with a more physically realistic age of $< 10$~Myr.

Figure \ref{fig_age_strata} shows the spatial distribution of $t<10$~Myr stars stratified by the ages obtained from the FLWO data; the LAICA data produce similar strata. The region is further divided into two sub-regions: the East sub-region near/around the O6 ionizing star HD~206267 is associated with the central stellar cluster Tr~37, while the West sub-region near/around the IC 1396A globule is expected to be associated mainly with stars that have been recently formed in the globule. Our data provide strong evidence that the stars in the East sub-region are systematically older than the stars in the West sub-region (Figure \ref{fig_age_strata}~(d)). Kolmogorov-Smirnov (K-S) and ``Fisher's exact'' tests give small probabilities that the two age samples are drawn from the same distribution\footnote{In the case of the FLWO ages, $P_{KS} = 1.6$\% and a $2\times2$ contingency table with younger/older stars in the East$:$West region $= 14/16 : 42/15$ is consistent with $P_{Fisher} = 1.8$\%. In the case of the LAICA ages, $P_{KS} = 0.01$\% and a $2\times2$ contingency table with younger/older stars in the East$:$West region $= 11/18 : 45/14$ is consistent with $P_{Fisher} = 0.08$\%. The derived probabilities are significantly lower than the conventional criterion for statistical significance of $\la 5$\% (http://en.wikipedia.org/wiki/Statistical\_significance). Thus the null hypothesis that the two samples are drawn from the same age distribution can be confidently rejected.}. The derived mean/median FLWO (LAICA) ages are 3.9~Myr/3.6~Myr and 2.8~Myr/2.2~Myr (4.4~Myr/4~Myr and 2.7~Myr/1.9~Myr) for the East and West sub-regions, respectively. These are consistent with the previously reported average age of $4$~Myr for the Tr~37 cluster (SA05). Typical age estimates for the stars in front of the globule are $t \la 2-3$~Myr. Meanwhile, younger ages ($t \la 1$~Myr) should be assumed for numerous disky objects that are embedded in the globule and lack optical counterparts (\S \ref{imf_analysis_globule_subsection}).

\section{SPATIAL STRUCTURE}\label{spatial_structure_section}

\begin{figure*}\footnotesize
\centering
\includegraphics[angle=0.,width=180mm]{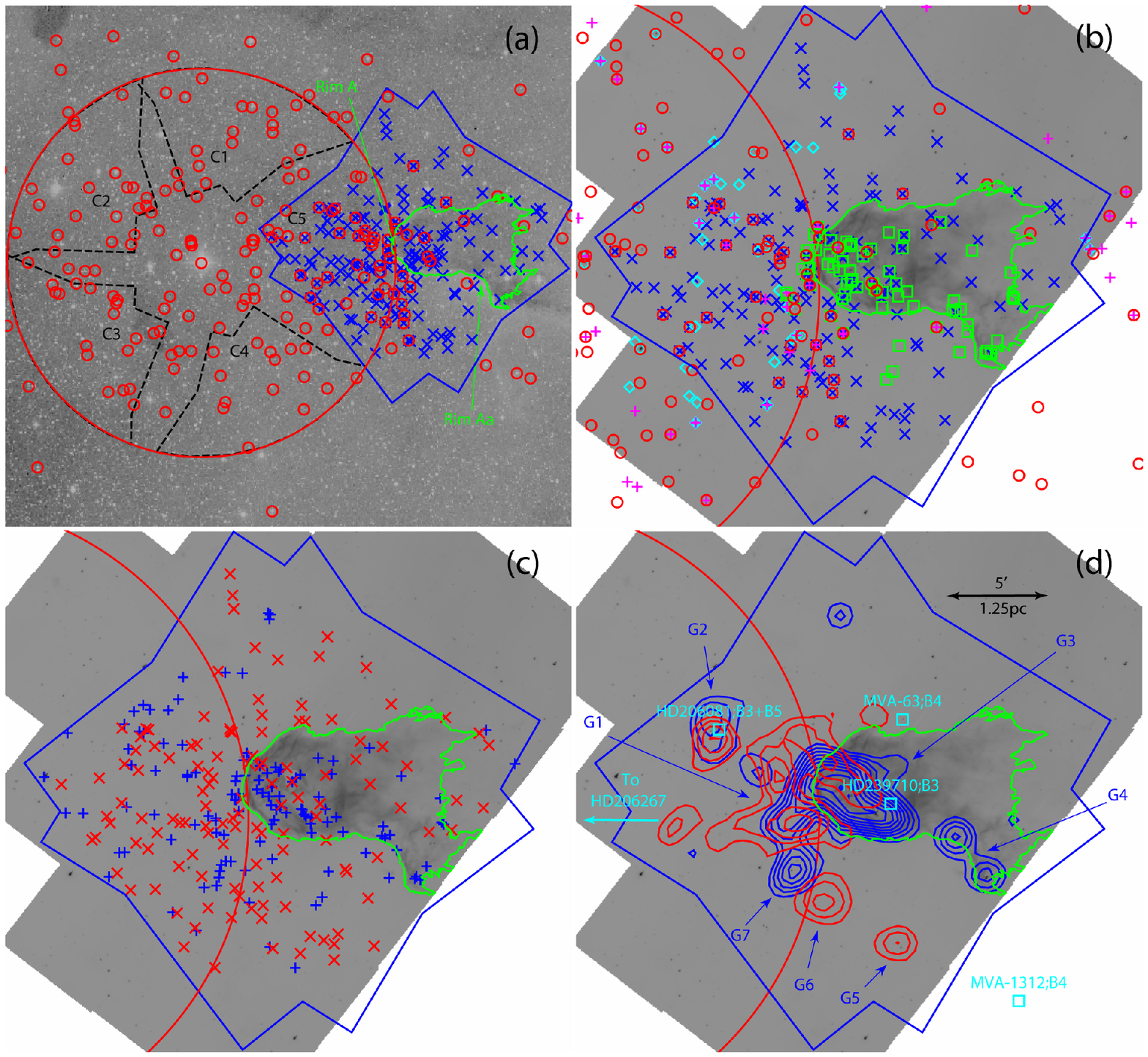}
\caption{Spatial distribution of known young members of the region over-plotted on DSS2-Red (a) and Spitzer-IRAC 8~$\mu$m (b-d) images. On all panels the big red circle outlines  the area of the central cluster Tr~37 of radius $R=15\arcmin$ centered on the O6 star HD~206267. The blue polygon outlines the ACIS field. The green polygon demarcates the globule, tracing $8.0$~$\mu$m emission from hot dust and PAHs. (a) Positions of young members from the SA05 (red $\circ$) and the X-ray (blue $\times$) catalogs. ``C1''through``C5'' label five equal-area segments at the edge of the central cluster. Positions of the two optically bright rims in front of the cloud are marked by arrows and labeled in green. (b) Positions of all known young members of the region within the ACIS field: optical SA05 (red $\circ$), X-ray selected (blue $\times$), MIR stars from MC09 (green $\sq$), H$\alpha$ stars from B11 (magenta $+$), and additional MIR selected stars from this study (cyan $\diamond$). (c) Separation on X-ray selected diskless stars (red $\times$) and disky stars (blue $+$) combined from MC09 and this work. (d) Contour maps of source surface densities from panel (c) smoothed with a Gaussian $\sigma = 1.4\arcmin$. Source density contours are in linear units from 0.75 to 1.5 sources per arcmin$^2$ for diskless (red), and from 0.75 to 3 sources per arcmin$^2$ for disky (blue) stars; both with a step of 0.25 sources per arcmin$^2$. Stellar enhancements above 1.25 source per arcmin$^{2}$ level are indicated by arrows and labeled in blue. Known YSOs with spectral type earlier than B6 are marked in cyan.  North is up, east is to the left. \label{fig_spatial_4panels}}
\end{figure*}

In this section we combine data from different stellar catalogs in order to examine spatial distributions of all known young stars in/around the globule and the neighboring Tr~37 cluster. For the central part of the Tr~37 cluster, we utilize optical data from SA05, while for region in and around the globule we consider the MIR catalog of MC09, the X-ray catalog, and our additional MIR catalog from Table \ref{tbl_nonXray_disky}. The latter three catalogs in turn include disky YSOs from \citet{Reach04} and \citet{Sicilia-Aguilar06} as well as all H$\alpha$ stars recently identified within the ACIS field by B11. See Figure \ref{fig_spatial_4panels} for the spatial distribution of these stars.

\subsection{Stellar surface density gradient from the SA05 sample}\label{spatial_subsection1}

Here we define the central portion of the Tr~37 cluster as a circular area of radius $R=15\arcmin$ (large red circle in Figure \ref{fig_spatial_4panels}) centered on the O6 star HD~206267 and bordering the head of the IC 1396A globule. We refer to this area hereafter as R15. Nearly the entire area is within the central 4-Shooter field (Figures 1, 9 in SA05), which has a uniform and deep optical exposure. The circular area comprises 150 optical young stars (Tables 1, 3, 5 in SA05). The cluster is not centrally concentrated. For instance, compared to the younger Orion Nebula Cluster (ONC) where the surface stellar density drops by a factor of $>25$ from the center to the $2.5$~pc radius \citep{Hillenbrand98}, the density in Tr~37 decreases by only a factor of two over a similar parsec scale.

Our goal here is to search for local stellar surface over-densities on the periphery of R15 area that could be related to triggered star formation in now-dissipated molecular cloudlets. The overlap region between R15 and the ACIS field (labeled as ``C5'' in Figure \ref{fig_spatial_4panels} (a)) is especially useful in regards to possible triggered star formation in the originally larger IC~1396A cloud. The periphery of R15 is divided by five control equal-area segments labeled as ``C1''--``C5''. From the SA05 master sample, we count 16, 20, 18, 15, and 35 stars in segments C1, C2, C3, C4, and C5, respectively. We thus find that the stellar surface density in C5 area is two times higher (with statistical significance $P > 99.5$\%) than the average density in the remaining periphery area of the central cluster. The $\sim 17$ extra stars could be associated with triggered star formation in the originally larger IC~1396A cloud.

\subsection{Increase in the census of YSOs in and around IC 1396A}\label{spatial_subsection2}

As X-ray emission from PMS stars is based on enhanced solar-type magnetic reconnection events rather than disk or accretion processes, X-ray selection delivers rich and clean samples of diskless stars missed by IR selection \citep{Feigelson10}. Previous studies have identified 129 YSO members within our ACIS field of view.  This study has doubled that census by identifying 130 additional stellar members in our X-ray (\S \ref{membership_subsection}) and MIR (\S \ref{non_chandra_members_subsection}) catalogs (Figure \ref{fig_spatial_4panels} (b)).  Twenty-nine (14 {\it Chandra} and 15 non-{\it Chandra}) exhibit an IR excess.

\subsection{Stellar surface density gradient from the combined stellar sample}\label{spatial_subsection3}

A map of the combined sample of 259 members is shown in Figure \ref{fig_spatial_4panels} (c).  They appear clustered rather than uniformly distributed, and the distributions of the disky (blue $+$) and diskless (red $\times$) stars appear to be offset horizontally from each other.  This is more plainly seen in the source surface density maps shown in panel (d), which were constructed using a fixed Gaussian kernel. 

Directly in front of the molecular globule, most of the diskless and a portion of the disky stars lie in a $>3$~pc long and $\sim 0.7$~pc wide layer. This result is robust against observation selection effects as the excess of the stars within the layer is seen for both the {\it Chandra} X-ray selected diskfree stars and the disky stars that were mainly identified through {\it Spitzer} IR selection. The position of the layer is also consistent with the positions of the very young optical stars (SA05, B11) and with the excess of the optical stars on the western border of Tr~37 that was discussed in \S \ref{spatial_subsection1}.

The layer can be further subdivided into a few stellar groupings: about 30 disky and diskless stars along the bright rim and to the east of the globule towards the direction of the major ionizing system HD~206267 (labeled as ``G1''); about 20 disky and diskless stars to the north-east of the globule around the B3+B5 high-mass system HD 206081 (``G2''); about 20 stars to the south-east of the globule (the disky ``G6'' and the diskless ``G7''). Some groupings are disky and others diskless, suggesting different ages from distinct RDI events in different cloud cores over millions of years (\S \ref{discussion_rdi_clumps_subsection}).

Inside the globule, the stellar population is dominated by disky objects, whose spatial distribution is also highly non-uniform and clumpy, with the bulk of the stars lying along the western edge of the globule that faces the ionizing system HD~206267. Two major stellar groupings are seen inside the cloud: a larger group composed of $\sim 50$ Class III/II/I sources at the head of the globule behind the bright optical Rim A (group ``G3''); and a smaller group composed of $\sim 10$ sources, mainly Class~I protostars, located behind optical Rim Aa in the portion of the cloud named SFO 36 (group ``G4'').

\subsection{Disk-fraction gradient}\label{spatial_subsection4}

The source surface densities of the diskless and disky stars lying within the ACIS field inside and in front of the cloud suggest a spatial gradient of apparent disk fraction (Figure \ref{fig_spatial_4panels} (d)). The highest disk fractions are inside the cloud, behind Rims A ($9/9 = 100$\%) and Aa ($35/47 = 75\pm6$\%)\footnote{Errors on disk fractions have been estimated using binomial distribution statistics as described by \citet{Burgasser03}.}. An intermediate disk fraction of $41/85 = 48\pm5$\% is found in the layer with the enhanced stellar surface density in front of the globule (groups ``G1'', ``G2'', ``G6'', and ``G7'' $=$G1267). Eastward of the layer, towards Tr~37, the disk fraction is lowest ($12/34=35\pm8$\%) and is consistent with a recent estimate of the intrinsic disk fraction \citep[39\%][]{Mercer09}. This disk-fraction gradient within our sample is a tentative finding, for two reasons. First, the gradient is statistically significant across the G1267-globule region, but not across the Tr37-G1267 region\footnote{For the G1267-globule area, a $2\times2$ contingency table with disky/diskfree ratio $= 41/44 : 35/12$ is inconsistent with the null hypothesis of no disk-fraction gradient ($P_{Fisher} = 0.5$\%, using Fisher's exact test). However, for the Tr37-G1267 area, a $2\times2$ contingency table with disky/diskfree ratio $= 12/22 : 41/44$ is consistent with the null hypothesis of no disk-fraction gradient ($P_{Fisher} = 23$\%.}. Second, the comparison is performed with apparent disk fractions, which are subject to a spatially varying X-ray sensitivity that is suppressed inside the globule.

\section{XLFs, IMFs FOR THE STELLAR POPULATIONS IN THE TR~37/IC~1396A REGION}\label{total_populations_section}

We construct X-ray luminosity functions (XLFs) and initial mass functions (IMFs) independently for the different projected regions on the sky. The separation into the older Tr~37 and younger (likely triggered) IC~1396A stellar populations will be made in \S \ref{implications_for_tsf_section}. For stars outside the globule in the ACIS field, two independent analysis are performed: an XLF analysis for purely X-ray selected stars (\S \ref{xlf_analysis_outside_globule_subsection}), and an IMF analysis for the combined X-ray and MIR YSO sample (\S \ref{imf_analysis_outside_globule_subsection}). The IMF analysis of the optical SA05 YSO sample is employed to measure the total stellar population within the $R=15\arcmin$ radius circular area around HD~206267 (\S \ref{imf_analysis_trumpler37_subsection}). Finally, the IMF of a stellar sample mainly composed of disky YSOs from MC09 combined with some additional X-ray stars is utilized to estimate the total number of stars inside the globule within the ACIS field (\S \ref{imf_analysis_globule_subsection}).

\subsection{XLF of the stellar population around the globule}\label{xlf_analysis_outside_globule_subsection}

Comparison of X-ray luminosity functions (XLFs) between newly studied young stellar clusters like IC 1396A and a well-characterized cluster like the ONC can give insight into any missing population of stars. The assumption of a universal XLF \citep{FeigelsonGetman05} has been used in the past to estimate total populations in several young clusters, such as Cep~B \citep{Getman06}, M17 \citep{Broos07},  NGC~6357 \citep{Wang07}, NGC~2244 \citep{Wang08}, NGC~2237 \citep{Wang10}, W40 \citep{Kuhn10}, Tr~15 \citep{Wang11}, Tr~16 \citep{Wolk11}, Sh~2-254/255/256/257/258 \citep{Mucciarelli11}, and NGC~1893 \citep{Caramazza11}. Notice that XLFs are relatively insensitive to age effects; between 0.1 and 10 Myr, only a slight change (roughly 0.3 in $\log L_X$) is found in the ONC stellar population \citep{Preibisch05,Prisinzano08}.

\begin{figure}
\centering
\includegraphics[angle=0.,width=85mm]{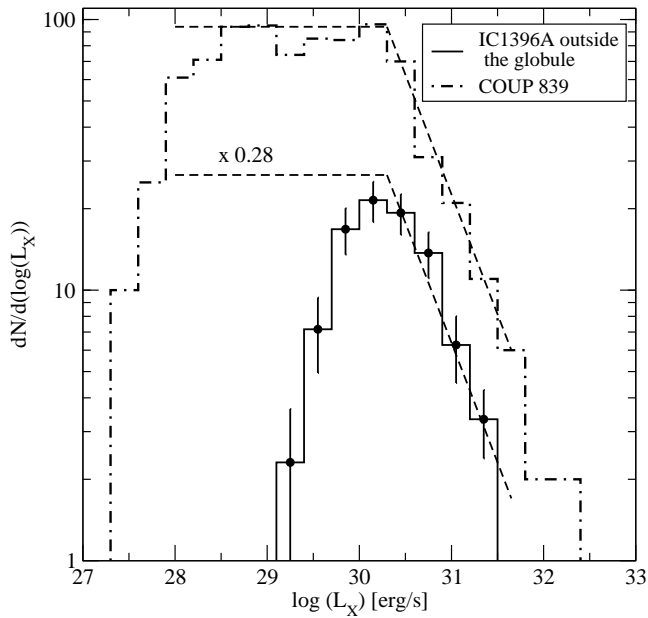}
\caption{Comparison between XLFs of the ONC and the population of the X-ray selected YSOs located outside the IC 1396A globule. The dashed-dotted histogram shows the COUP unobscured cool sample of 839 stars; the solid histogram is for the sample of 91 X-ray YSOs with available X-ray luminosity information located around the IC 1396A globule. Error bars indicate 68\% confidence intervals (1~$\sigma$) from Monte Carlo simulated distributions as described in the text. The thin dashed lines are added to aid the eye; they are based on the shape of the ONC XLF and scaled downward to match the IC 1396A XLF. \label{fig_xlf}}
\end{figure}

For the X-ray selected young stars related to IC 1396A, X-ray luminosities ($L_X$) and absorbing column densities ($N_H$) are generated using the non-parametric method {\it XPHOT} \citep{Getman10}. The concept of the method is similar to the long-standing use of color-magnitude diagrams in optical and infrared astronomy, with X-ray median energy replacing color index and X-ray source counts replacing magnitude. Empirical X-ray spectral templates derived from bright sources from the {\it Chandra} Orion Ultradeep Project \citep[COUP;][]{Getman05} are further used to translate apparent photometric properties of weak PMS stars into their intrinsic properties. The advantage of the {\it XPHOT} method over a traditional parametric spectral modeling (e.g.,  {\it XSPEC}) is that it is more accurate for very faint sources and provides both statistical and systematic (due to uncertainty in X-ray model) errors on derived intrinsic fluxes and absorptions. The {\it XPHOT} results are given in Table \ref{tbl_derived_props}.

Figure \ref{fig_xlf} shows the XLF for all X-ray selected YSOs with available $L_X$ information that lie outside the IC 1396A globule. The error bars on the XLF were generated from 1000 Monte-Carlo simulated XLF distributions with individual X-ray luminosities randomly drawn from Gaussian distributions with mean equal to the measured source's $\log(L_X)$ and variances based on the corresponding statistical and systematic errors summed in quadrature. Over-plotted is a template XLF for the COUP unobscured population \citep[839 cool ONC stars in][]{Feigelson05}, which is nearly complete down to $M = 0.1$~M$_{\odot}$ \citep[\S7.2 in][]{Getman06}. 

The IC 1396A XLF falls quickly below $\log(L_X) < 30.0$~erg~s$^{-1}$ due to incompleteness, with a conservative value for the completeness limit at around $\log(L_X) = 30.5$~erg~s$^{-1}$. From the $M$-$L_X$ correlation of \citet{Preibisch05b,Telleschi07} this corresponds to a mass value of $M \simeq 1$~M$_{\odot}$\footnote{The alternative IMF analysis (\S \ref{imf_analysis_outside_globule_subsection}) will further show that the inclusion of additional non-{\it Chandra} Class~II members selected by IR-excess will push the mass completeness limits for all known YSOs outside the globule down to $0.3-0.4$~M$_{\odot}$.}. Matching the two cluster XLFs in the range $30.5 < \log(L_X) < 31.5$~erg~s$^{-1}$ requires scaling down the ONC unobscured population by a factor of $\sim 0.28$, yielding a total of $\sim 235$ stars down to $0.1$~M$_{\odot}$ outside the IC 1396A globule within the ACIS FOV. Notice that the observed weak X-ray sources without available $L_X$ information do not affect the results of the analysis because their X-ray luminosities are expected to be below the completeness limit.

\subsection{IMF of the stellar population around the globule}\label{imf_analysis_outside_globule_subsection}

In the area outside the globule covered by the ACIS field and all four bands of the IRAC mosaic, we examine the stellar IMF separately for diskless and disky stars. To compensate for the effect of higher X-ray detection efficiency of Class~III stars, we use the disky stellar sample that is a combination of the {\it Chandra} and non-{\it Chandra} IR-excess member young stars (Tables \ref{tbl_derived_props} and \ref{tbl_nonXray_disky} in this work, and Table~2 from MC09) while retaining the diskless stellar sample as purely composed of X-ray stars (Table \ref{tbl_derived_props}). 

Approximate stellar mass estimates are obtained based on star locations in the $J$ versus $J-H$ color-magnitude diagram and theoretical stellar model tracks derived by \citet{Baraffe98} (for $0.02\, M_{\odot} \leqslant M \leqslant 1.4\, M_{\odot}$) and \citet{Siess00} (for $1.4\, M_{\odot} \leqslant M \leqslant 7.0\, M_{\odot}$). These color-magnitude diagrams are shown in Figures \ref{fig_JvsJH_acis} and \ref{fig_JvsJH_disky}. We are aware that these mass estimates are subject to significant uncertainties and may be incompatible with individual masses obtained by other methods such as optical spectroscopy \citep[][their Appendix section]{Kuhn10}.  However, here we are not so much interested in exact masses for individual objects as in the general trends of mass distributions and consistent treatment for comparison between different stellar populations.

\begin{figure}
\centering
\includegraphics[angle=0.,width=85mm]{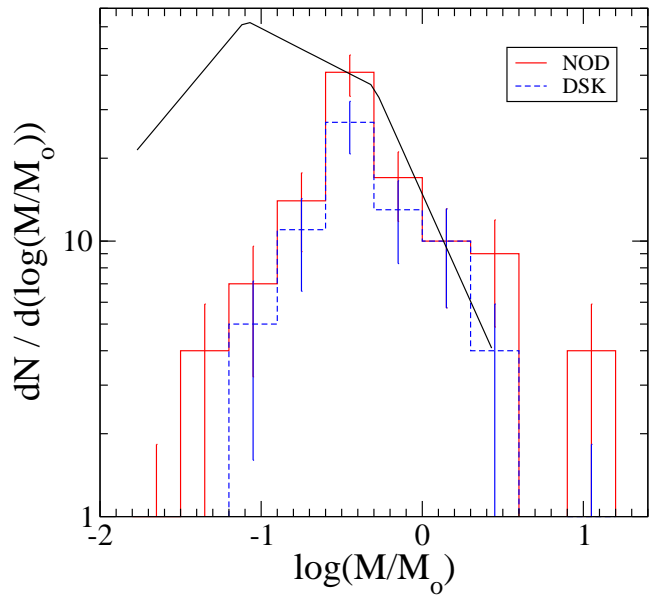}
\caption{The mass distribution of all known X-ray and IR-excess selected YSOs outside the globule within the ACIS field of view, assuming an age of $\sim 2$~Myr. Red and blue histograms are for diskless and disky YSOs, respectively. The $1\sigma$ Poisson errors \citep{Gehrels86} are also shown. The black line represents the scaled version of the Galactic field IMF \citep{Kroupa02}. \label{fig_imf_outside_globule}}
\end{figure}

Typical derived age estimates for the stars in front of the globule are $t \la 2-3$~Myr (\S \ref{age_analysis_section}). Figure \ref{fig_imf_outside_globule} shows the IMF of YSOs outside the globule for an assumed age of 2~Myr. The mass distribution within errors follows the shape of the Galactic field IMF with rough mass completeness limits at $0.3-0.4$~M$_{\odot}$. The disk fraction obtained by counting diskless and diskbearing stars with masses above the completeness limit is $41 \pm 5$\%. The inferred total number of stars is around $250$ stars down to $0.1$~M$_{\odot}$. Similar results are obtained from the IMF analysis with the assumption of an age of $3$~Myr (figure is not shown). The total number of stars ($\sim 250$) obtained here from the IMF is consistent with the total ($\sim 235$) derived from the age-insensitive XLF analysis (\S \ref{xlf_analysis_outside_globule_subsection}).

\subsection{IMF of the central part of Tr~37}\label{imf_analysis_trumpler37_subsection}

We estimate the total number of young stars in the central part of Tr~37 within the $R=15\arcmin$ circular area centered on the O6 star HD~206267 using the optical catalog of SA05. For the central cluster we assume the age of $\sim 4$~Myr (SA05, and \S \ref{age_analysis_section} here). As in \S \ref{imf_analysis_outside_globule_subsection}, stellar mass estimates are derived from the $J$ versus $J-H$ color-magnitude diagram.

\begin{figure}
\centering
\includegraphics[angle=0.,width=85mm]{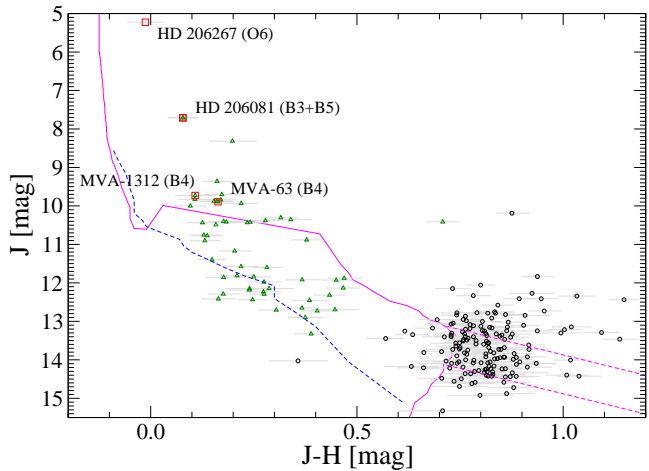}
\caption{The NIR color-magnitude diagram for all (inside and outside the ACIS field) optical YSOs from SA05. Low-mass stars are shown as black circles; intermediate and high-mass stars are shown as green triangles. The unabsorbed zero-age main sequence and 4~Myr PMS isochrone at $870$~pc distance are shown as dashed blue and solid magenta lines, respectively. Reddening vectors of $A_V = 5$~mag for masses $0.5$ and $1$~M$_{\odot}$ are indicated by dashed magenta lines. Stars of spectral type earlier than B6 are outlined by red $\sq$ and labeled. \label{fig_JvsJH_sa05}}
\end{figure}

Figure \ref{fig_JvsJH_sa05} shows that about 20 intermediate-mass YSO candidates lie far to the left from the 4~Myr isochrone with some of them along the ZAMS locus. It is unclear if these are members of an old distributed population or older stars unrelated to the region. These stars are excluded from the IMF analysis.

\begin{figure}
\vspace*{0.2in}
\centering
\includegraphics[angle=0.,width=85mm]{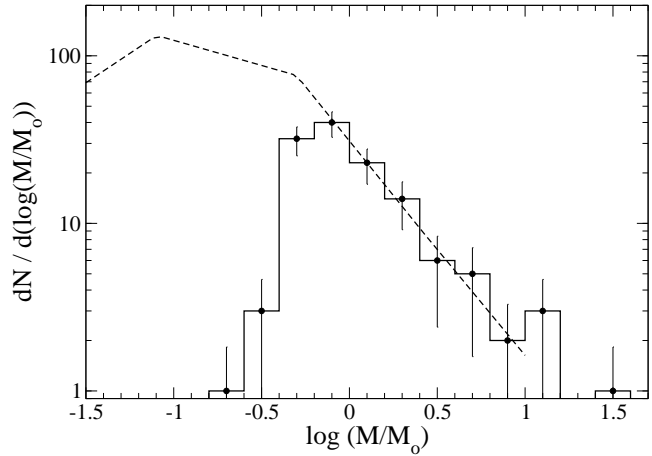}
\caption{The mass distribution for optical SA05 stars located within the circular area of radius $R=15\arcmin$ around the O6 star HD~206267 (solid histogram). The dashed line represents the scaled version of the Galactic field IMF \citep{Kroupa02}. The $1\sigma$ Poisson errors \citep{Gehrels86} are also shown. \label{fig_imf_central_cluster}}
\end{figure}

Figure \ref{fig_imf_central_cluster} shows that the mass distribution for the optical stars in the central cluster within errors follows the shape of the Galactic field IMF. The approximate mass completeness limit for the optical sample is $0.8$~M$_{\odot}$. The inferred total number of stars in the circular $R = 15\arcmin$ cluster area is found to be 480 stars down to $0.1$~M$_{\odot}$. This IMF population and the XLF population from \S \ref{xlf_analysis_outside_globule_subsection} are consistent in the overlap region between the circular $R = 15\arcmin$ cluster area and the ACIS field.

\subsection{Estimate of the stellar population inside the globule}\label{imf_analysis_globule_subsection}

The $J$ versus $J-H$ diagram of X-ray selected YSOs shown in Figure \ref{fig_JvsJH_acis} suggests that the X-ray detection efficiency of disky stars is lower than that of diskless stars. The $J$ versus $J-H$ diagram for all known disky YSOs with available NIR information (Figure \ref{fig_JvsJH_disky}) further shows that {\it Chandra} did not detect many low-mass and/or highly obscured disky stars identified by means of {\it Spitzer} data. For the analysis of the stellar population inside the globule, a region where {\it Chandra}'s sensitivity deteriorates, we rely heavily on the {\it Spitzer} catalog of disky stars from MC09.

There are 18 X-ray selected diskless stars and 45 IR-excess selected and/or X-ray selected diskbearing stars projected against the globule (Table \ref{tbl_derived_props} here and Table 2 from MC09). With the assumption that all of these observed stars are globule members, the apparent disk fraction averaged across the entire globule is $45/(18+45)$ or $70$\%. Twenty of the diskbearing stars have been classified as Class~I objects (MC09). The radiative transfer model fits to the optical-IR-radio band data of a mid-IR selected sample of YSOs in the globule yield ages of $\la 0.2$~Myr and $\la 1$~Myr for the Class I and Class II objects, respectively \citep{Reach09}. Considering the high apparent disk fraction, the high fraction of Class~I objects, and the young ages derived for the selected sample of Class I/II YSOs by \citet{Reach09}, we reasonably assume an age $\la 1$~Myr for the entire population in the globule. 

\begin{figure}
\centering
\includegraphics[angle=0.,width=85mm]{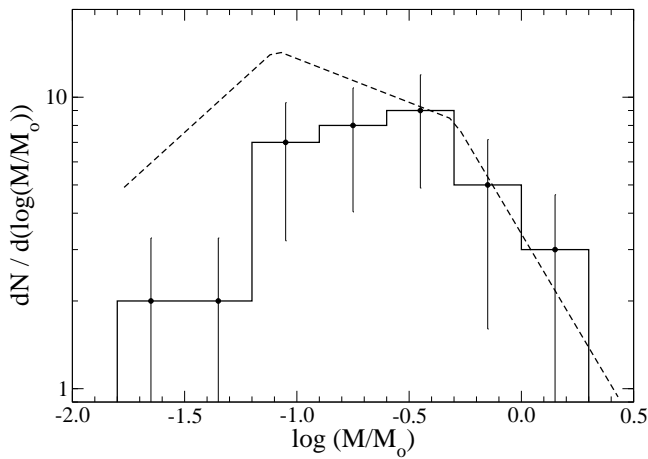}
\caption{The mass distribution of disky stars found within the globule. Mass distribution of thirty eight young disky stars (36 from MC09 and additional 2 new from the X-ray catalog) with available mass estimates, which lie projected against the globule, is shown as a solid histogram. The dashed line represents the scaled version of the Galactic field IMF \citep{Kroupa02}. The $1\sigma$ Poisson errors \citep{Gehrels86} are also shown. \label{fig_imf_inside_globule}}
\end{figure}

The IMF analysis here is restricted to the diskbearing population in the globule. These are 38 out of 45 observed disky objects with available stellar mass estimates obtained from the $J$ versus $J-H$ color-magnitude diagram (Figure \ref{fig_JvsJH_disky}). Their mass distribution is shown in Figure \ref{fig_imf_inside_globule}. Despite the low counting statistics, the shape of the distribution is reminiscent of that of the Galactic field IMF. The IMF analysis gives a crude number on total population of diskbearing stars in the globule of $\sim 40$ stars down to $0.1$~M$_{\odot}$. This is likely a low limit because 7 diskbearing stars without $JHK$ information have not been included in the analysis. 

With respect to the diskless population in the globule, 18 X-ray selected dikless stars projected against the globule have not been included in the IMF analysis. Out of these 18 X-ray stars, at least 10 have relatively high source extinctions, comparable to those of the diskbearing population in the globule (Figure \ref{fig_JvsJH_acis}). Assuming an age of $1$~Myr, all 10 would have masses above $0.1$~M$_{\odot}$. Thus, the total number of disky and diskless stars in the part of the globule covered by the ACIS field of view is $\ga 60$.

\section{IMPLICATIONS FOR TRIGGERED STAR FORMATION}\label{implications_for_tsf_section}

\subsection{Observational evidence for RDI triggered star formation}\label{discussion_rdi_evidences_subsection}

Spatial-age gradients found in stellar populations in/around BRCs are often viewed as strong observational evidence for triggered star formation. The youngest stars are typically embedded at the edge of the cloud facing the ionizing star, and older stars are aligned within and in front of the bright rim towards the ionizing star. For example, spatio-temporal gradients and triggered star formation have been reported in BRCs and pillars on the edges of large H\,{\sc{ii}} regions in the Eagle Nebula \citep{Sugitani02}, IC 1848 \citep{Matsuyanagi06, Ogura07}, IC 1396 \citep{Getman07,Ikeda08,Choudhury10}, Sharpless 155 \citep{Getman09}, Sharpless 171 and Sharpless 296 \citep{Chauhan09}, and the Carina Nebula \citep{Smith10}.

In \S \ref{age_analysis_section} and \ref{spatial_structure_section} we establish the spatio-temporal gradient of young stars from the center of the IC~1396A cloud toward the HD 206267 O6.5f star. The identified stellar members are not uniformly distributed around the globule but rather are concentrated along the ionized rim, both outside (stellar groups ``G1''$+$``G2''$+$``G6''$+$``G7'' in Figure \ref{fig_spatial_4panels}) and inside (``G3''$+$``G4'') the cloud. The clustering of young stars in front of the cloud is also consistent with the previously reported arc-shaped layer of $\sim 20$ low-mass high accretion stars from B11 and/or a dozen of $\sim 1$~Myr optical stars from SA05. The stars located inside the cloud have typical ages of $\la 1$~Myr (\S \ref{imf_analysis_globule_subsection}), the stars found right in front of the cloud have typical ages of $\la 2-3$~Myr (\S \ref{age_analysis_section}), and the stars located further to the east, in the Tr~37 cluster, are older with ages of $\ga 3.5$~Myr (\S \ref{age_analysis_section}). This age gradient is also consistent with our tentative finding that the apparent disk fraction is decreasing from $75-100$\% inside the cloud cores, to $48$\% in front of the cloud, and to $35$\% further to the east towards HD~206267 (\S \ref{spatial_structure_section}).

The results from this work, combined with the results from the previous studies, thus provide strong evidence that the Tr~37/IC~1396A region possesses many of the observational features of an RDI triggered star formation: \begin{enumerate}

\item The presence of an exciting star (HD~206267) and a bright-rimmed cloud (IC 1396A) protruding inside the H\,{\sc{ii}} region and surrounded by an ionized rim (Rim~A and Rim~Aa) facing the exciting star (Figure \ref{fig_intro}).

\item The presence of dense molecular cores close to the rim \citep[][their Figure 8]{Patel95} and \citep[][their Figure 14a]{Weikard96}. This is consistent with the scenario where the expanding ionization front from HD~206267 penetrates the inter-core cloud region and drives shocks into the cores, compressing them sufficiently to trigger gravitational collapse. 

\item A rough equilibrium between the H\,{\sc{ii}} pressure and the cloud pressure found at the interfaces between the Rim~A and the head of the IC~1396A globule \citep{Reach09} and the Rim~Aa and the SFO~36 cloud \citep{Morgan04}. This indicates that the cloud could have been recently shocked by a photoionization shock. In addition, for the SFO~36 cloud, \citet{Morgan09,Morgan10} measure relatively high values of CO excitation temperature and turbulent velocity dispersion, which are consistent with those of other shocked cloud candidates.

\item A clumpy spatial distribution of young stars both inside and in front of the cloud (Figure \ref{fig_spatial_4panels}) suggesting recent formation in distinct molecular cores. The proximity of the stars to the ionized rim immediately suggests a physical relationship between star formation and the ionization front propagating through the globule, and can be explained within the framework of the RDI model. 

\item The spatio-temporal gradient of young stars, both age and disk-gradient, oriented towards the exciting star (Figures~\ref{fig_age_strata}-\ref{fig_spatial_4panels} here,  Figure 11 in SA05, and Figure 18a in B11).

\end{enumerate}

\subsection{Contribution of triggered star formation to the total population of the IC 1396 H\,{\sc{ii}} region}\label{discussion_tsf_contribution_subsection}

\subsubsection{Motivation}\label{discussion_motivation_subsection}

The overall effect of triggered star formation on the global star formation rate of the Galaxy is difficult to ascertain \citep{Elmegreen11}. A handful of studies attempted to estimate the effect. From an observational side, counting IRAS point sources embedded in BRCs of numerous H\,{\sc{ii}} regions, \citet{Sugitani91} estimate the contribution of triggered star formation from expanding H\,{\sc{ii}} regions to be only $\sim 5$\% by mass of the stars in the Galaxy. \citet{Deharveng10} find that $26$\% of the mid-IR shells from the {\it Spitzer}-GLIMPSE survey harbor UCH\,{\sc{ii}}s and/or methanol masers, likely indicating triggering of massive star formation. From comparison of positions of very young and massive stars identified in the RMS survey in respect to locations of numerous mid-IR shells from the {\it Spitzer}-GLIMPSE survey, \citet{Thompson11} estimate the fraction of Galactic massive stars triggered by expanding H\,{\sc{ii}} regions to be $14-30$\%.

From a theoretical perspective, by turning on and off an irradiation feedback on a turbulent $10^{4}$~M$_{\odot}$ molecular cloud, \citet{Dale07b} find that triggered star formation might increase the total amount of star formation by as much as $30$\%. However, their SPH simulations show no observable characteristics for distinguishing triggered from spontaneously formed stars. Although it is often difficult to make such a distinction in observational studies of individual H\,{\sc{ii}} regions,  in \S \ref{discussion_rdi_evidences_subsection} and \ref{discussion_centralpart_subsection} we argue that an RDI triggered stellar population can be identified in the case of the Tr~37/IC~1396 A region. Hence, the contribution of triggered star formation in the central part of the IC 1396 H\,{\sc{ii}} region, and with some additional assumptions in the entire H\,{\sc{ii}} region, can be estimated.

\subsubsection{Contribution in the central part of the H\,{\sc{ii}} region}\label{discussion_centralpart_subsection}

In \S \ref{total_populations_section} we establish the total populations of stars down to $0.1$~M$_{\odot}$ for different areas of the region: $\sim 480$ stars within the $R=15\arcmin$ radius circular area around HD~206267; $\sim 235-250$ stars in the area outside the globule covered by the ACIS field; and $\ga 60$ YSOs inside the globule within the ACIS field. These numbers are employed here to estimate the input of triggered star formation to the total population of the central part of the region.

Based on the arguments presented in \S \ref{discussion_rdi_evidences_subsection}, it is reasonable to propose that most of the YSOs inside the IC 1396A cloud and the bulk of the stars in the stellar groups ``G1''$+$ ``G2'' $+$ ``G6'' $+$ ``G7'' (=G1267) in front of the cloud (Figure \ref{fig_spatial_4panels}) were formed through the RDI process. The notion is also consistent with our finding of a 2-fold increase in the surface density of the optical SA05 stars at the western border of the central Tr~37 cluster (\S \ref{spatial_structure_section}). Assuming that the 2-fold increase is fully due to a triggered population, we obtain $\sim 20$\% as a conservative estimate on the fraction of the old stellar members of Tr~37  within the G1267 group, which implies that $\sim 50$ out of the 60 X-ray YSOs detected in G1267 have a triggered origin.

The total number of triggered stars in G1267 group in front of the BRC is estimated to be $\sim 80 = 50 \times (235/146)$ stars down to 0.1~$M_{\odot}$, where $\sim 50$ is the number of {\it Chandra} observed triggered candidates in G1267, 146 is the number of {\it Chandra} observed stars in the entire ACIS field outside the globule, and 235 is the inferred total number of stars in the entire ACIS field outside the globule. This number combined with the conservative (low limit) estimate on the total stellar population inside the globule ($\ga 60$ stars; \S \ref{imf_analysis_globule_subsection}) gives the total triggered population in IC 1396A of $\ga 140$ stars down to $0.1$~M$_{\odot}$. This is likely a lower limit as we do not include possible contribution of the sparse group of stars with a high apparent disk fraction seen right in front of the Rim Aa (Figure \ref{fig_spatial_4panels}c). 

In contrast, the total population of the central Tr~37 cluster is $480 - 35/2/150 \times 480 = 424$ stars down to $0.1$~M$_{\odot}$, where $480$ is an estimate on the total stellar population within the $R=15\arcmin$ radius circular area around HD~206267, $35$ and $150$ are numbers of the observed optical SA05 stars located  within the overlap area ``C5'' and the entire $R=15\arcmin$ circular area, respectively, and $2$ accounts for the increase in the surface density of the stars within the overlap area due to the presence of the triggered population. The contribution of the triggered stellar population to the total population of the central part of the IC 1396 H\,{\sc{ii}} region to date is thus $\ga 25$\% ($\ga 140 / (424+140)$).

\subsubsection{Contribution for the entire H\,{\sc{ii}} region}\label{discussion_entire_subsection}

A rough estimate of the contribution of triggered star formation for the entire IC~1396 H\,{\sc{ii}} region can be further obtained by extrapolating our results for the triggered population in/around IC 1396A cloud combined with the information on the low-mass high-accretion stars from the survey of the entire IC 1396 H\,{\sc{ii}} region by B11. This requires the following additional assumptions. First, we assume that the total number of stars in the entire IC~1396 H\,{\sc{ii}} region is similar to that of the Orion Nebula Cluster. This is supported by the fact that they each have equal numbers of stars B9 or earlier, and thus should have roughly the same total stellar populations\footnote{From Table 1 of \citet{Stelzer05}, the ONC has 9 stars between B3-O7 and 5 stars between B9-B4. From Table 5 of SA05 and adding HD 206267 and HD 206183, IC 1396 has 3 probable members between B3-O6 and 13 between B9-B4. So they each have $\sim 15$ stars B9 or earlier. Furthemore, the most massive systems in both regions have similar spectral types --- O6 star HD 206267 and O9.5 star HD 206183 in IC 1396 $vs.$ O6-7 star $\theta^{1}$~Ori~C and O9.5 star $\theta^{2}$~Ori~A in ONC.}. Here we assume that the ONC population is 2800 stars \citep{Hillenbrand98}. Second, we assume that the birth of all low-mass high-accretion B11 stars, which are found in clusters in front of the BRCs, was triggered. This is true for the ACIS field, where out of 22 B11 stars 21 are likely associated with the triggered population: 19 belong to the G1267 stellar group in/around Rim A, one is within the cloud's head, and the other one is right in front of the Rim Aa. Based on Figure 9 from B11 we count $>15$ YSOs clustered in front of Rim E, about 10 stars in front of Rim B, several high-accretion stars in front of Rims J and H, and $>10$ stars in front of some other, unlabeled, BRCs, totaling $>40$ observed B11 triggered stars outside the ACIS field. Third, we assume that the ratio of the number of the observed B11 triggered stars to the total number of the associated triggered stars, $R_t$, is constant across the H\,{\sc{ii}} region. Within the ACIS field, $R_t = 21 / (>140) \sim 15$\%. The contribution of triggered star formation for the entire H\,{\sc{ii}} region is then $\ga 14$\% ($(>40+21)/R_t/2800$).

Combining this with the estimate for the central cluster in \S \ref{discussion_centralpart_subsection}, $\ga 14-25$\% of the stars in the IC 1396 region have a triggered origin. This is consistent with the various observational and theoretical estimates reviewed in \S \ref{discussion_motivation_subsection}.

However, it has to be stressed that we are observing IC 1396 and other giant  H\,{\sc{ii}} regions at only a single moment in their histories. IC 1396 has expanded to a radius $~12$ parsecs over few million years, and the  H\,{\sc{ii}} shock likely triggered star formation in many now-dissipated molecular cloud cores. The  H\,{\sc{ii}} region is likely to persist for several million years into the future with supernova remnant shocks in addition to ultraviolet radiation and OB wind shocks. These shocks will encounter more cores that are now embedded in the larger cloud. Thus, if the cluster can trigger $14-25$\% of its population from the cloud cores that are active today, it seems plausible that it can trigger much more over the full lifetime of the HII region.

\subsection{Other aspects of triggered star formation}\label{discussion_other_aspectsoftsf_subsection}

\subsubsection{Star Formation Efficiency}\label{discussion_sfe_subsection}

Measured star formation efficiencies (SFEs) of entire giant molecular clouds typically vary from a few to several percent \citep[][and references therein]{Evans09}, whereas SFEs of embedded clusters associated with individual massive dense molecular cores are significantly higher ranging between 10\% and 30\% \citep[][and references therein]{Lada03}.

For the total triggered population in/around IC 1396A of $>140$ stars, assuming an average mass of $\sim 0.5$~$M_{\odot}$, consistent with a standard IMF, and adding individual masses of likely triggered high-mass B-type stars in the region, $\ga 10$~M$_{\odot}$ for the B3$+$B5 binary star HD 206081 ({\it Chandra} source \#382) and several~M$_{\odot}$ for the B3 star HD 239710 ({\it Chandra} source \# 135; Figure \ref{fig_spatial_4panels}d), we obtain an estimated total triggered stellar mass of $M_{star} \sim 90$~M$_{\odot}$. The mass of the observed molecular gas in IC 1396A is only $M_{gas} \sim 200$~M$_{\odot}$ today \citep{Patel95, Weikard96}. The inferred current SFE for the IC 1396A stellar population is thus $M_{star}/(M_{star}+M_{gas}) \sim 30$\% at the top of typical SFE range measured in other active star forming regions. 

Even higer SFEs of $>40$\% were recently reported in a number of cases of likely triggered star formation, e.g., Cep~B/OB3b \citep{Getman09}, Sh 2-233IR \citep{Yan10}, L935/NGC~7000 \citep{Toujima11}. However, the gas mass detected today could be only a fraction of the original cloud mass, so the SFE would be lower with respect to the original gas mass. Thus, the high values of apparent SFE in IC 1396A and the other regions of triggered star formation indicate either an efficient star formation or  an efficient removal of molecular gas by photoevaporation and stellar winds.

\subsubsection{Long-lived RDI triggering in a clumpy cloud}\label{discussion_rdi_clumps_subsection}

Intermediate-mass stars are often born surrounded by small groups of a few to a few dozen lower-mass PMS stars \citep{Testi99,Adams01}. The stellar group ``G2'' is composed of $\ga 15$ YSOs around the B3$+$B5 binary system HD 206081 (Figure \ref{fig_spatial_4panels}c,d). The high apparent disk fraction of $\sim 60$\%, the presence of 6 high-accretion YSO systems from B11, and the filamentary structure seen at 24$\mu$m coincident with the stellar group suggest that the group was formed very recently. The projected distance between ``G2'' and the stellar group ``G1'', located right along the optical Rim A, is $1.5$~pc. Theoretically, it is possible that stars recently formed in ``G1'' having acquired velocity dispersion $\ga 1$~km~s$^{-1}$ have drifted $1-2$~pc in all directions, populating the field around ``G1'' including the location of ``G2'' with triggered stars. However, the distinct appearance of the two groups, and the notion of simultaneous formation of intermediate-mass stars and PMS siblings, favor the view that the star drift does not play an important role here and that the present locations of the two groups are directly associated with locations of their distinct parental molecular cores. This is analogous to the current state of the IC 1396A cloud, where the two distinct stellar groups ``G3'' and ``G4'' inside the cloud are each associated with its own molecular core.

This implies continuing star formation over millions of years when the RDI mechanism occurs in a larger than the presently seen IC 1396A cloud composed of multiple clumps. This view is consistent with the recent global picture of an  H\,{\sc{ii}} region expanding  into a non-uniform clumpy molecular cloud, where triggered star formation would take place through simultaneous enhancement of density and global radiation-driven implosion of numerous pre-existing molecular clumps \citep{Walch11}.

\section{CONCLUSIONS}\label{conclusions_section}

Signatures of small-scale triggered star formation on the periphery of H\,{\sc{ii}} regions can be erased on short timescales and thus, it is often difficult to identify sites of triggered star formation and to measure the impact of triggering processes. There are only a handful of studies in the literature that have attempted to estimate the contribution of triggered star formation from expanding H\,{\sc{ii}} regions. Infrared emitting protostars in cloud cores and larger populations of PMS stars with spatial gradients in stellar age indicative of long-standing triggering processes have both been seen associated with several clouds on the periphery of expanding H\,{\sc{ii}} regions. Much of the progress is due to the fruitful combination of X-ray and optical/infrared surveys; the former captures the older diskless PMS population while the latter detects the disky and accretion population.

IC 1396 is a nearby, large shell-like H\,{\sc{ii}} region, where traces of recent triggered star formation are still evident. We present new {\it Chandra} X-ray data and auxiliary IR {\it Spitzer} and optical FLWO/LAICA data of YSOs associated with the central cluster Trumpler~37 and the adjacent bright-rimmed cloud IC 1396A. These data are merged with the data from previously published optical-IR stellar catalogs of the region. The goal of this work is to identify, understand, and quantify the triggered stellar population in the region associated with the IC 1396A cloud and to estimate the contribution of triggered star formation to the entire H\,{\sc{ii}} region.

Out of $415$ X-ray sources detected in the Tr~37/IC~1396A area, 175 are identified as YSOs and classified on 124 diskless and 51 disky objects (\S \ref{ir_optical_counterparts_section}). The majority of the remaining X-ray sources are contaminants, mainly extragalactic sources unrelated to the region (\S \ref{membership_subsection}). In addition to the 175 X-ray emitting YSOs, we identify 42 non-$Chandra$ IR-excess low-mass members of the region (\S \ref{non_chandra_members_subsection}). Combining these data with the previous optical (SA05, B11) and IR (MC09) catalogs of YSOs yields a total of 259 YSOs in/around IC~1396A cloud within the ACIS field, half of which are new members discovered in this work (\S \ref{spatial_structure_section}).

The age and spatial distributions of young stars reveal a spatio-temporal gradient of stars from the IC~1396A cloud toward the primary ionizing star HD 206267 (\S \ref{age_analysis_section} and \ref{spatial_structure_section}). Young stars are found to be clustered along the ionized rim, both outside and inside the cloud. The clustering of young stars in front of the cloud is consistent with the previously reported arc-shaped layer of the low-mass high accretion stars (B11) and/or the $\sim 1$~Myr old optical stars (SA05). Quantitatively, the number of stars clustered in front of the cloud is consistent with the 2-fold increase in the stellar surface density at the western border of the central Tr~37 cluster (\S \ref{spatial_structure_section} and \ref{discussion_centralpart_subsection}). The stellar age increases from $\la 1$~Myr inside the cloud, to $<2-3$~Myr in front of the cloud, to $\sim 4$~Myr towards the central Tr~37 cluster. Combined with the results from the previous studies, these findings provide strong evidence for RDI triggered star formation over millions of years in the Tr~37/IC~1396A region (\S \ref{discussion_rdi_evidences_subsection}). Based on this evidence, we believe that the majority of the YSOs found inside the cloud and the majority of the stars found clustered in front of the cloud were likely formed through the RDI process extending over $2-3$~Myr.

By scaling the XLF/IMF measurements obtained for different YSO samples in the region (\S \ref{total_populations_section}) to the number of the observed triggered stars, we estimate a total of $>140$ triggered stars down to 0.1~M$_{\odot}$ in/around the IC~1396A cloud, constituting $\ga 25$\% of the total popualtion in the central part of the region today (\S \ref{discussion_centralpart_subsection}). Furthermore, with a few additional assumptions we estimate the contribution of the triggered stellar population to the total population of the entire IC 1396 H\,{\sc{ii}} region to be $>14$\% (\S \ref{discussion_entire_subsection}). Thus, the currently active clouds on the periphery of IC 1396 add at least $14-25$\% to the current cluster population. When past and future clouds are considered, the contribution of triggering to the star formation in the molecular cloud could be much higher.

The inferred apparent value of star formation efficiency for the triggered population in IC~1396A is $30\%$ using current gas cloud masses (\S \ref{discussion_sfe_subsection}). Our findings support a picture of continuing star formation over millions of years when the RDI mechanism occurs in a larger than the presently seen IC 1396A cloud composed of multiple clumps (\S \ref{discussion_rdi_clumps_subsection}). 


\section*{Acknowledgments}

We thank the anonymous referee for helpful comments. We thank Leisa Townsley (Penn State) for her role and financial support in the development of the wide variety of {\it Chandra}-ACIS data reduction and analysis techniques and tools used in this study. We thank Kevin Luhman (Penn State) for assistance with {\it Spitzer}-IRAC data analysis. We thank Robert Gutermuth (University of Massachusetts) for development of {\it Spitzer} data analysis tools. We benefited from scientific discussions with Matt Povich and Leisa Townsley (Penn State). This work is supported by the {\it Chandra} ACIS Team (G. Garmire, PI) through the SAO grant SV4-74018 and by the {\it Chandra} GO grant SAO \# GO0-11027X (K. Getman, PI). A.S.A. acknowledges support from the Spanish ``Ramon y Cajal'' program. The Center for Exoplanets and Habitable Worlds is supported by the Pennsylvania State University, the Eberly College of Science, and the Pennsylvania Space Grant Consortium. This work is based on observations made with the {\it Spitzer Space Telescope}, which is operated by the Jet Propulsion Laboratory, California Institute of Technology under a contract with NASA. We also use data products of the Two Micron All Sky Survey, which is a joint project of the University of Massachusetts and the Infrared Processing and Analysis Center/California Institute of Technology, funded by NASA and NSF.

\bsp

\label{lastpage}

\end{document}